\documentclass[twocolumn]{pasj01}

\RequirePackage{mathpazo}
\RequirePackage[T1]{fontenc} 

\usepackage{color}
\usepackage{comment}

\def\commenta{$^*$}
\def\commentb{$^\dagger$}
\def\commentc{$^\ddagger$}

\def\commente{$^\|$}

\newcounter{author}
\def\authorcount#1#2{\refstepcounter{author}\label{#1}
                     \altaffiltext{\ref{#1}}{#2}}

\Received{$\langle$reception date$\rangle$}
\Accepted{$\langle$acception date$\rangle$}
\Published{$\langle$publication date$\rangle$}

\begin{document}
\SetRunningHead{Y. Tampo et al.}{dwarf nova superoutbursts observed by Seimei telescope and VSNET}

\title{Spectroscopic and Photometric Observations of Dwarf Nova Superoutbursts by the 3.8 m Telescope Seimei and the Variable Star Network }

\author{
        Yusuke~\textsc{Tampo}\altaffilmark{\ref{affil:Kyoto}*},
        Keisuke~\textsc{Isogai}\altaffilmark{\ref{affil:Kyoto}}$^,$\altaffilmark{\ref{affil:KyotoOkayama}}$^,$\altaffilmark{\ref{affil:MuSCAT}},
        Naoto~\textsc{Kojiguchi}\altaffilmark{\ref{affil:Kyoto}},
        Hiroyuki~\textsc{Maehara}\altaffilmark{\ref{affil:NAOJOkayama}}$^,$\altaffilmark{\ref{affil:KyotoOkayama}}$^,$\altaffilmark{\ref{affil:mhh}}, 
        Kenta~\textsc{Taguchi}\altaffilmark{\ref{affil:Kyoto}},
        Taichi~\textsc{Kato}\altaffilmark{\ref{affil:Kyoto}},
        Mariko~\textsc{Kimura}\altaffilmark{\ref{affil:riken}}$^,$\altaffilmark{\ref{affil:Kyoto}},
        Yasuyuki~\textsc{Wakamatsu}\altaffilmark{\ref{affil:Kyoto}},
        Masaaki~\textsc{Shibata}\altaffilmark{\ref{affil:Kyoto}},
        Daisaku~\textsc{Nogami}\altaffilmark{\ref{affil:Kyoto}},
        Miho~\textsc{Kawabata}\altaffilmark{\ref{affil:Kyoto}}$^,$\altaffilmark{\ref{affil:KyotoOkayama}},
        Keiichi~\textsc{Maeda}\altaffilmark{\ref{affil:Kyoto}},
        Kosuke~\textsc{Namekata}\altaffilmark{\ref{affil:NAOJ}}$^,$\altaffilmark{\ref{affil:Kyoto}},
        Soshi~\textsc{Okamoto}\altaffilmark{\ref{affil:Kyoto}},
        Masaaki~\textsc{Otsuka}\altaffilmark{\ref{affil:Kyoto}}$^,$\altaffilmark{\ref{affil:KyotoOkayama}},
        Burgaz~\textsc{Umut}\altaffilmark{\ref{affil:Kyoto}}$^,$\altaffilmark{\ref{affil:umut}},
        Shumpei~\textsc{Nagoshi}\altaffilmark{\ref{affil:Kyoto}},
        Hiroshi~\textsc{Itoh}\altaffilmark{\ref{affil:Ioh}}, 
        Tonny~\textsc{Vanmunster}\altaffilmark{\ref{affil:Van1}}$^,$\altaffilmark{\ref{affil:Van2}}, 
        Tam\'as~\textsc{Tordai}\altaffilmark{\ref{affil:Polaris}}, 
        Geoffrey~\textsc{Stone}\altaffilmark{\ref{affil:sge}}, 
        Katsura~\textsc{Matsumoto}\altaffilmark{\ref{affil:oku}}, 
        Daiti~\textsc{Fujii}\altaffilmark{\ref{affil:oku}}, 
        Ryota~\textsc{Matsumura}\altaffilmark{\ref{affil:oku}}, 
        Momoka~\textsc{Nakagawa}\altaffilmark{\ref{affil:oku}}, 
        Nodoka~\textsc{Takeuchi}\altaffilmark{\ref{affil:oku}}, 
        Yuki~\textsc{Zenkou}\altaffilmark{\ref{affil:oku}}, 
        Elena~P.~\textsc{Pavlenko}\altaffilmark{\ref{affil:CrAO}}, 
        Kirill~A.~\textsc{Antonyuk}\altaffilmark{\ref{affil:CrAO}}, 
        Nikolaj~V.~\textsc{Pit}\altaffilmark{\ref{affil:CrAO}}, 
        Oksana~I.~\textsc{Antonyuk}\altaffilmark{\ref{affil:CrAO}}, 
        Julia~V.~\textsc{Babina}\altaffilmark{\ref{affil:CrAO}}, 
        Aleksei~V.~\textsc{Baklanov}\altaffilmark{\ref{affil:CrAO}}, 
        Aleksei~A.~\textsc{Sosnovskij}\altaffilmark{\ref{affil:CrAO}}, 
        Sergey~Yu.~\textsc{Shugarov}\altaffilmark{\ref{affil:shu2}}$^,$\altaffilmark{\ref{affil:saimsu}}, 
        Nataly~\textsc{Katysheva}\altaffilmark{\ref{affil:saimsu}},
        Drahom\'ir~\textsc{Chochol}\altaffilmark{\ref{affil:shu2}},
        Shawn~\textsc{Dvorak}\altaffilmark{\ref{affil:dks}}, 
        Pavol~A.~\textsc{Dubovsky}\altaffilmark{\ref{affil:Dubovsky}}, 
        Tom\'a\v{s}~\textsc{Medulka}\altaffilmark{\ref{affil:Dubovsky}}, 
        Igor~\textsc{Kudzej}\altaffilmark{\ref{affil:Dubovsky}}, 
        Seiichiro~\textsc{Kiyota}\altaffilmark{\ref{affil:kis}}, 
        Alexandra~M.~\textsc{Zubareva}\altaffilmark{\ref{affil:saimsu}}$^,$\altaffilmark{\ref{affil:zub}}, 
        Alexandr~A.~\textsc{Belinski}\altaffilmark{\ref{affil:saimsu}}, 
        Natalia~P.~\textsc{Ikonnikova}\altaffilmark{\ref{affil:saimsu}},
        Marina~A.~\textsc{Burlak}\altaffilmark{\ref{affil:saimsu}}, 
        Yasuo~\textsc{Sano}\altaffilmark{\ref{affil:san}}$^,$\altaffilmark{\ref{affil:san2}},
        Masanori~\textsc{Mizutani}\altaffilmark{\ref{affil:mzm}}, 
        Javier~\textsc{Ruiz}\altaffilmark{\ref{affil:rui1}}$^,$\altaffilmark{\ref{affil:rui2}}$^,$\altaffilmark{\ref{affil:rui3}}, 
        Roger~\textsc{D.Pickard}\altaffilmark{\ref{affil:rpc}}, 
        Franz-Josef,~\textsc{Hambsch}\altaffilmark{\ref{affil:ham1}}$^,$\altaffilmark{\ref{affil:ham2}}$^,$\altaffilmark{\ref{affil:dfs}}, 
        Sjoerd~\textsc{Dufoer}\altaffilmark{\ref{affil:dfs}}, 
        Stephen~M.~\textsc{Brincat}\altaffilmark{\ref{affil:Brincat}},
        Charles~\textsc{Galdies}\altaffilmark{\ref{affil:gch}}, 
        Kenneth~\textsc{Menzies}\altaffilmark{\ref{affil:mzk}}, 
        Masayuki~\textsc{Moriyama}\altaffilmark{\ref{affil:myy}},
        Mitsutaka~\textsc{Hiraga}\altaffilmark{\ref{affil:hrm}}, 
        Yutaka~\textsc{MAEDA}\altaffilmark{\ref{affil:mdy}}, 
        Kenji~\textsc{Hirosawa}\altaffilmark{\ref{affil:hsk}},
        Masao~\textsc{Funada}\altaffilmark{\ref{affil:fnm}}, 
        and Minoru~\textsc{Yamamoto}\altaffilmark{\ref{affil:ymo}} 
}

\authorcount{affil:Kyoto}{
    Department of Astronomy, Graduate School of Science, Kyoto University, Oiwakecho,
    Kitashirakawa, Sakyo-ku, Kyoto 606-8502}
\email{$^*$tampo@kusastro.kyoto-u.ac.jp}

\authorcount{affil:KyotoOkayama}{
     Okayama Observatory, Kyoto University, 3037-5 Honjo, Kamogatacho,
     Asakuchi, Okayama 719-0232, Japan}

\authorcount{affil:MuSCAT}{
    Department of Multi-Disciplinary Sciences, Graduate School of Arts and Sciences, 
    The University of Tokyo, 3-8-1 Komaba, Meguro, Tokyo 153-8902, Japan}

\authorcount{affil:NAOJOkayama}{
     Subaru Telescope Okayama Branch Office, National Astronomical Observatory of Japan, 
     National Institutes of Natural Sciences, 3037-5 Honjo, Kamogata, Asakuchi, Okayama 719-0232, Japan}

\authorcount{affil:mhh}{Variable Star Observers League in Japan (VSOLJ), Okayama, Japan}

\authorcount{affil:riken}{
     Extreme Natural Phenomena RIKEN Hakubi Research Team, Cluster for Pioneering Research, 
     RIKEN, 2-1 Hirosawa, Wako, Saitama 351-0198, Japan}

\authorcount{affil:NAOJ}{
    National Astronomical Observatory of Japan, National Institutes of Natural Sciences, 
    2-21-1 Osawa, Mitaka, Tokyo 181-8588, Japan}

\authorcount{affil:umut}{
     Department of Astronomy and Space Sciences, University of Ege, 35100, \.Izmir, Turkey}

\authorcount{affil:Ioh}{
     Variable Star Observers League in Japan (VSOLJ),
     1001-105 Nishiterakata, Hachioji, Tokyo 192-0153, Japan}
     
\authorcount{affil:Van1}{
     Center for Backyard Astrophycis Belgium, 
     Walhostraat 1a, B-3401 Landen, Belgium}

\authorcount{affil:Van2}{
    Center for Backyard Astrophycis Extremadura, 
    e-EyE Astronomical Complex, 
    ES-06340 Fregenal de la Sierra, Spain}

\authorcount{affil:Polaris}{
     Polaris Observatory, Hungarian Astronomical Association,
     Laborc utca 2/c, 1037 Budapest, Hungary}

\authorcount{affil:sge}{
    Center for Backyard Astrophysics Sierras, 44325 Alder Heights Road, Auberry, CA 93602, USA}

\authorcount{affil:oku}{
Osaka Kyoiku University, 4-698-1 Asahigaoka, Osaka 582-8582, Japan}

\authorcount{affil:CrAO}{ 
     Federal State Budget Scientific Institution Crimean Astrophysical
     Observatory of RAS, Nauchny, 298409, Republic of Crimea}
     
\authorcount{affil:shu2}{
Astronomical Institute of the Slovak Academy of Sciences, 05960 Tatransk\'a Lomnica, Slovakia}

\authorcount{affil:saimsu}{
    Sternberg Astronomical Institute, Lomonosov Moscow State University, Universitetsky Ave.,13, Moscow 119992, Russia}

\authorcount{affil:dks}{
Rolling Hills Observatory, 1643 Nightfall Drive, Clermont, Florida 34711, USA}

\authorcount{affil:Dubovsky}{
     Vihorlat Observatory, Mierova 4, 06601 Humenne, Slovakia}
     
\authorcount{affil:kis}{
Variable Star Observers League in Japan (VSOLJ), 7-1 Kitahatsutomi, Kamagaya, Chiba 273-0126, Japan}

\authorcount{affil:zub}{
    Institute of Astronomy, Russian Academy of Sciences, Moscow 119017, Russia}
    
\authorcount{affil:san}{
    Variable Star Observers League in Japan (VSOLJ), Nishi juni-jou minami 3-1-5, Nayoro, Hokkaido, Japan}

\authorcount{affil:san2}{
    Observation and Data Center for Cosmosciences, Faculty of Science, Hokkaido University, Kita-ku, Sapporo, Hokkaido 060-0810, Japan}

\authorcount{affil:mzm}{
    Variable Star Observers League in Japan (VSOLJ),
    Okayama, Japan}

\authorcount{affil:rui1}{
    Observatorio de Cantabria, Ctra. de Rocamundo s/n, Valderredible, 39220, Cantabria, Spain.}
\authorcount{affil:rui2}{
    Instituto de Fisica de Cantabria (CSIC-UC), Avda. Los Castros s/n, 39005, Santander, Spain}
\authorcount{affil:rui3}{
    Agrupacion Astronomica Cantabra, Apartado 573, 39080, Santander, Spain}
    
\authorcount{affil:rpc}{
    The British Astronomical Association, Variable Star Section (BAA VSS), Burlington House, Piccadilly, London, W1J 0DU, UK}

\authorcount{affil:ham1}{
Groupe Européen d’Observations Stellaires (GEOS), 23 Parc de Levesville, 28300 Bailleau l’Evêque, France}
\authorcount{affil:ham2}{
Bundesdeutsche Arbeitsgemeinschaft für Veränderliche Sterne (BAV), Munsterdamm 90, 12169 Berlin, Germany}

\authorcount{affil:dfs}{
Vereniging Voor Sterrenkunde (VVS), Oostmeers 122 C, 8000 Brugge, Belgium}

\authorcount{affil:Brincat}{
     Flarestar Observatory, San Gwann SGN 3160, Malta}

\authorcount{affil:gch}{
    Institute of Earth Systems, University of Malta, Malta}

\authorcount{affil:mzk}{
    Center for Backyard Astrophysics (Framingham), 318A Potter Road, Framingham, MA 01701, USA}

\authorcount{affil:myy}{
Variable Star Observers League in Japan (VSOLJ),
290-383, Ogata-cho, Sasebo, Nagasaki, Japan}

\authorcount{affil:hrm}{Variable Star Observers League in Japan (VSOLJ), Shimonoseki, Yamaguchi, Japan}

\authorcount{affil:mdy}{
Variable Star Observers League in Japan (VSOLJ), Kaminishiyamamachi 12-2, Nagasaki, Nagasaki, Japan}

\authorcount{affil:hsk}{
Variable Star Observers League in Japan (VSOLJ), 216-4 Maeda, Inazawa-cho, Inazawa-shi, Aichi, Japan}

\authorcount{affil:fnm}{Variable Star Observers League in Japan (VSOLJ), Chiba, Japan}
\authorcount{affil:ymo}{Variable Star Observers League in Japan (VSOLJ), Aichi, Japan}


\KeyWords{accretion, accretion disk --- novae, cataclysmic variables --- stars: dwarf novae}

\maketitle

\begin{abstract}
We present spectroscopic and photometric observations of 17 dwarf-nova superoutbursts obtained by KOOLS-IFU mounted on the 3.8 m telescope Seimei at Okayama Observatory of Kyoto University and through VSNET collaboration.
Our spectroscopic observations for six outbursts were performed within 1 d from their optical peak.
11 objects (TCP J00590972+3438357. ASASSN-19ado, TCP J06073081-0101501, ZTF20aavnpug, ASASSN-19ady, MASTER OT J061642.05+435617.9, TCP J20034647+1335125, ASASSN-20kv, ASASSN-20kw, MASTER OT J213908.79+161240.2, and ASASSN-20mf)  
were previously unknown systems, and our observations enabled quick classification of their transient type.
These results illustrate that Seimei telescope has the capability to conduct quick follow-up observations of unknown transients.
Our photometric observations yielded that 11 objects are WZ Sge-type dwarf novae and their candidates, and the other six objects are SU UMa-type dwarf novae and their candidates.
The He II 4686\AA~ emission line was clearly detected among ASASSN-19ado, TCP J06073081-0101501 and MASTER OT J213908.79+161240.2, whose association with a spiral arm structure in an accretion disk has been suggested in the previous studies. 
Our result suggests that a higher-inclination system shows a stronger emission line of He II 4686\AA, as well as larger-amplitude early superhumps.

\end{abstract}

\section{Introduction}
\label{sec:1}
Cataclysmic variables (CVs) are a close binary system composed of a primary white dwarf (WD) and a secondary low-mass star. 
The secondary star fills up its Roche lobe, transferring mass into the primary Roche lobe through the inner Lagrangian point $L_1$. 
Dwarf novae (DNe) are a subclass of CVs which possess an accretion disk and show recurrent outbursts (for a review of CVs and DNe, see \cite{war95book,hel01book}).

U Gem-type DNe show normal outbursts, producing typically a 2-5-mag brightening for a few days.
The mechanism of normal outbursts is understood as the thermal instability in an accretion disk (\cite{osa74DNmodel}, \cite{mey81DNoutburst}).
SU UMa-type DNe, one subclass of DNe, are characterised by superoutbursts. 
Superoutbursts are generally longer and brighter than normal outbursts.
Superhumps are observed during a superoutburst, which are 0.1 - 0.5 mag variations with the period a few percent longer than the orbital one. 
Superoutbursts are explained by the thermal - tidal instability model \citep{whi88tidal, osa89suuma, hir90SHexcess}. 
WZ Sge-type DNe form a subclass of SU UMa-type DNe, showing mainly superoutbursts every 5 - 30 yr, and seldom normal outbursts due to their low mass-transfer rates \citep{osa02wzsgehump, kat15wzsge}. 
WZ Sge-type DNe are characterised by early superhumps, and part of WZ Sge-type DNe shows rebrightenings.
Early superhump is a double-wave modulation observed in an early phase (1 - 10 d) of a WZ Sge-type DN superoutburst  \citep{kat15wzsge}. 
It is widely known that the period of an early superhump is almost identical with the orbital one within less than $\le 0.1 \%$ difference \citep{ish02wzsgeproc, kat02wzsgeESH}.
Rebrightening is an outburst following a main superoutburst before returning to quiescence, whose origin is still poorly understood \citep{ric92wzsgedip, kat98super, ham00DNirradiation, osa01egcnc, mey15suumareb}.

\citet{cla84sscyg, hes84sscyg} studied the evolution of optical spectra from quiescence to whole outburst cycle in SS Cyg.
Quiescence flat spectra, dominated by  Balmer and some He I emission lines, are replaced by bluer continuum and narrower absorption lines during an outburst (see examples in \cite{mor02DNspectralatlas}).
This is because, during a DN outburst, the entire accretion disk becomes optically thick and the accretion disk outshines the whole system \citep{hor85zcha}.
More specifically, due to foreshortening and limb-darkening effects, the strength of absorption strongly depends on the inclination of a system; a higher inclination system tends to show stronger emission components (\cite{mar90ippeg, war95book} and references there in).
In fact, highly inclined and eclipsing IP Peg does not show absorption lines even during its outbursts \citep{pic89ippeg}.
In Z Cha, \citet{hon88zcha} noted that the difference of optical spectra between normal outbursts and superoutbursts is the strong enhancement of He II 4686 \AA~ and C III / N III Bowen blend in superoutbursts, which would reflects the difference of outburst energetic.

The spectral evolution during WZ Sge-type DN superoutbursts has been well studied in WZ Sge \citep{bab02wzsgeletter, nog04wzsgespec}, GW Lib \citep{hir09gwlib, vanspa10gwlib}, EZ Lyn \citep{iso15ezlyn}, and SSS J122221.7-311525 \citep{neu17j1222}.
However, due to the long outburst cycles and a small number of known WZ Sge-type DNe, systematic analyses and comparisons of the spectra during WZ Sge-type DN outbursts have been a challenge.
In the superoutburst spectra of WZ Sge-type DNe, along with  Balmer and He I series, high excitation lines such as He II 4686\AA~ and  C III / N III Bowen blend tend to be observed more frequently than in normal outbursts (\cite{bab02wzsgeletter, nog04wzsgespec, hir09gwlib, kat15wzsge} and references there in).
Especially on He II 4686\AA~ emission line, \citet{bab02wzsgeletter, kuu02wzsge} found the possible association between the early superhumps and the spiral arm structure in the accretion disk, which was obtained through the Doppler tomography of He II 4686 \AA.
The presence of a spiral arm during a DN outburst is also observed in SS Cyg \citep{ste01spiralwave}, U Gem \citep{gro01ugemspiral}, and IP Peg \citep{har99ippeg} through similar analyses.
Therefore the emission line of He II 4686 \AA~ can be a good tracer of a spiral structure in an accretion disk (e.g. \citep{zho20lamostDN}).
However, the emission mechanism of these high excitation lines is still controversial.
Since WZ Sge-type DNe are a low mass-transfer rate system, these high excitation lines are not likely associated with a hot spot.
\citet{mor02DNspectralatlas} suggested two scenarios for the formation of He II 4686\AA; either a spiral structure itself or irradiated temperature-inversion layer on the spiral structure.
Observational results from WZ Sge and GW Lib preferred the irradiation scenario based on the line profile evolution \citep{nog04wzsgespec, hir09gwlib}.
In general, these high excitation lines are present only in the early stage of superoutbursts.
On the other hand, sodium doublet (Na D, $\lambda$ 5890 / 5896 \AA) is observed even during the post-superoutburst stage as absorption (EG Cnc; \cite{pat98egcnc}) or emission (WZ Sge; \cite{nog04wzsgespec}, GW Lib; \cite{vanspa10gwlib}, SSS J122221.7-31152; \cite{neu17j1222}, V3101 Cyg; \cite{tam20j2104}).
These authors have discussed possible association of Na D line with the mass reservoir, which can be a mass supplier during a rebrightening phase \citep{kat98super, hel01book}.
\citet{nog04wzsgespec} discussed that the origins of Na D line at the optical peak and in the post-superoutburst stage are likely different.

In this paper, Section \ref{sec:2} describes an overview of observations.
Section \ref{sec:3} presents our results of spectroscopic and photometric observations. 
We discuss the characteristics of our spectra, light curve profiles and correlation between the observational features and system parameters in Section \ref{sec:4}.
We give our summary in Section \ref{sec:5}.

\section{Observations and Analysis}
\label{sec:2}

\begin{longtable}{cp{1.3cm}p{1.5cm}p{1.5cm}p{1.5cm}c}
  \caption{Log of spectroscopic observations}\label{tab:1}
  \hline              
    Name & DNe  & $P_{\rm orb}$ & BJD  & Days   &  Peak - Quiescence \\ 
     & Type\commenta &  &  Observed &  since peak  & magnitude  \\ 
        &   & (d) & & (d) & mag (filter)\\
    \hline
    \hline
\endfirsthead
    \multicolumn{6}{l}{\commenta UGWZ: WZ Sge-type, UGSU: SU UMa-type, ':' mark means a candidate of that type.}\\
    \multicolumn{6}{l}{\commentb The superhump period is listed instead of $P_{\rm orb}$.}\\
\endfoot
    AL Com                          & UGWZ      & 0.056668              & 2458591.2     & 2.5       & 12.9($CV$) - 19.8($V$) \\   
                                    &           &                       & 2458592.1     & 3.4       &  \\   
                                    &           &                       & 2458594.2     & 5.5       &  \\   
    EQ Lyn                          & UGWZ      & 0.0528(1)             & 2458757.3     & 3.3       & 10.3($CV$) - 19.1($g$) \\
    MASTER OT J234843.23+250250.4   & UGSU      & 0.032\commentb        & 2458793.2     & 4.5       & 14.4($CR$) - 20.3($r$) \\   
    NN Cam                          & UGSU      & 0.07391(2)            & 2458826.1     & 8.9       & 12.6($CV$) - 18.5($V$) \\   
    TCP J00590972+3438357           & UGWZ      & 0.0543(1)             & 2458831.1     & 0.2       & 12.5($CV$) - 21.0($g$) \\   
    ASASSN-19ado                    & UGWZ      & 0.06318(5)            & 2458842.0     & 2.3       & 12.8($g$) - 23:($g$) \\   
    ASASSN-19ady                    & UGSU:     & ---                   & 2458845.1     & 1.3       & 14.7($g$) - 21.5($r$) \\   
    TCP J06073081-0101501           & UGWZ      & 0.062(1)              & 2458877.1     & 0.5       & 13.9($CV$) - $>$22 \\   
    DDE35                           & UGSU      & 0.132(2)\commentb     & 2458929.2     & ---       & 15.2($CV$) - 19.3($G$) \\   
                                    &           &                       & 2458930.1     & ---       &  \\   
    MASTER OT J061642.05+435617.9   & UGWZ      & 0.0549(1)             & 2458929.1     & 1.7       & 15.0($CR$) - $>$23 \\   
    ZTF20aavnpug                    & UGWZ:     & ---                   & 2458968.2     & 6.2       & 15.3($G$) - $>$23 \\   
    PQ And                          & UGWZ      & 0.0559(2)             & 2459003.3     & $>$4.1    & 10.0($CV$) - 19.2($V$) \\   
    TCP J20034647+1335125           & UGWZ      & 0.05537(4)            & 2459077.2       & 0.2       & 12.6($CR$) - 21.5($r$) \\
    ASASSN-20kv                     & UGWZ:     & 0.05604(2)\commentb   & 2459088.2       & 0.4       & 14.1($g$)  - 23:($g$)  \\
    ASASSN-20kw                     & UGSU:     & ---                   & 2459089.3       & 0.2       & 14.7($g$)  - 23:($g$) \\
    MASTER OT J213908.79+161240.2   & UGWZ:     & 0.0584(1)             & 2459119.2       & 0.6       & 15.5($CR$) - $>$23 \\
    ASASSN-20mf                     & UGSU:     & ---                   & 2459119.2       & 4.3       & 14.5($g$)  -  $>$23 \\
\hline
\end{longtable}

In this paper, we present the observations of 17 DN outbursts.
The low-resolution spectroscopic observations in this paper were carried out using the Kyoto Okayama Optical Low-dispersion Spectrograph with an Integral Field Unit (KOOLS-IFU; \cite{mat19koolsifu}) mounted on the 3.8-m telescope Seimei  at Okayama Observatory, Kyoto University \citep{kur20seimei}.
The log of our spectroscopic observations is listed in Table \ref{tab:1}.
The wavelength coverage of VPH-Blue of KOOLS-IFU is 4200 - 8000 \AA~, and its wavelength resolution is $R = \lambda/\Delta\lambda~\sim$~400 - 600. 
The data reduction was performed using IRAF in the standard manner (bias subtraction, flat fielding, aperture determination, spectral extraction, wavelength calibration with arc lamps and normalization by the continuum).
All the observation epochs are described in Barycentric Julian Date (BJD).

Our time-resolved CCD photometric observations were made with the Variable Star Network collaborations (VSNET; \cite{VSNET}). 
The logs of photometric observations are listed in Table E2 - E18 \footnote{Table E2 - E18 is available only on the online edition as Supporting Information. }. 
The zero-point of all the VSNET data were adjusted to the data of the $V$ or $CV$ band observations for each object.
We also use the photometric data from ASAS-SN Sky Patrol \citep{ASASSN, koc17ASASSNLC}, the Zwicky Transient Facility (ZTF; \cite{ZTF}) Public Data Release 4, ZTF alert broker Lasair \citep{lasair} and $\it{Gaia}$ Photometric Science Alerts \citep{wyr12gaiaalert} to examine the global light curve profiles, especially on the type of rebrightenings and on the outburst cycles.
The phase dispersion minimization (PDM) method was used for period analysis \citep{PDM}. 
The 90$\%$ confidence range of $\theta$ statistics by the PDM method was determined following \citet{fer89error, Pdot2}. 
Before period analysis, global trend of the light curve was removed by subtracting a smoothed light curve obtained by locally weighted polynomial regression (LOWESS; \cite{LOWESS}).

\section{Results}
\label{sec:3}

\begin{longtable}{cccccccc}
  \caption{Equivalent Width (EW) and Full Width at Half Maximum (FWHM) of our spectra}\label{tab:2}
  \hline              
    \multicolumn{2}{c}{} & \multicolumn{2}{c}{H$\alpha$} & \multicolumn{2}{c}{H$\beta$} & \multicolumn{2}{c}{H$\gamma$}\\
    Name & BJD Observed & EW\commente & FWHM & EW\commente & FWHM & EW\commente & FWHM \\ 
     &  &  & (\AA) &  & (\AA) &  & (\AA) \\ 
    \hline
    \hline
\endfirsthead
    \multicolumn{8}{l}{\commente A negative EW for an emission line.} \\
\endfoot
    AL Com                          & 2458591.2 & ---       & ---       & $-$7.3(2)     & 62.5(9)   & ---       &--- \\
                                    & 2458592.1 & ---       & ---       & $-$3.7(1)     & 35.3(8)   & --- & --- \\
                                    & 2458594.2 & ---       & ---       & $-$4.4(2)       & 81.67(9)   & --- & --- \\
    EQ Lyn                          & 2458757.3 & $-$ 6.6(6)  & 24.1(6) & $-$4.70(8)  & 28.0(7)   & $-$4.6(1)  & 20.5(5) \\
                                    & 2458757.3 &   1.3(2)  & 17.2(5) & ---     &   ---     &   --- &   --- \\
    MASTER J2348                    & 2458793.2 & ---        & ---       &  7.8(1)  & 47.6(8)   & ---       & ---       \\
    NN Cam                          & 2458826.1 & $-$7.7(1)  & 20.6(4)  & $-$12.4(2)  & 31.1(4)   & $-$9.7(1) & 19.5(2) \\
    TCP J0059                       & 2458831.1 & $-$2.7(1)   & 29.3(9) & $-$3.26(9)  & 22.7(8)   & $-$1.81(7)  & 15.2(5) \\
    ASASSN-19ado                    & 2458842.0 & ---         & ---     &  $-$3.5(2)    &  31.1(6)   & $-$2.7(1)       & 28.3(5)   \\
                                    & 2458842.0 & 20.51(7)    & 11.80(4)    &  1.58(6)    &  10.6(3)   & 0.57(4)       & 8.1(5)   \\
    ASASSN-19ady                    & 2458845.1 & ---       & ---       & $-$3.4(1)     & 18.4(5)     & -4.64(7)      & 19.7(8)   \\
    TCP J0607                       & 2458877.1 &  2.52(6)   &  9.1(3)      & $-$2.9(1)   & 38.7(9)   & 2.01(8)       & 25.4(8)   \\
    DDE35                           & 2458929.2 & 46.1(2)    & 24.13(6)  & 26.2(1)   & 21.82(9)     & 20.8(1) & 22.9(2) \\
                                    & 2458930.1 & 39.1(1)  & 25.29(8) & 17.2(1) &  20.4(1)   & 44.3(1)       & 37.9(1)       \\
    MASTER J0616                    & 2458929.1 & ---       & ---       & $-$3.2(1)     & 31.0(9)   & ---       & ---   \\
    ZTF20aavnpug                    & 2458968.2 & 22.05(7)   & 14.12(5)  & ---          & ---           & ---       & ---   \\
    PQ And                          & 2459003.3 & 5.67(7)  & 12.2(1)  & $-$3.8(1)     & 36.1(9)   & $-$6.2(1) & 32.6(7) \\
    TCP J2003                       & 2459077.2 & $-$3.33(9) & 28.7(8)  & $-$3.33(7)    & 21.6(9)   & $-$3.6(1) & 21.3(9) \\
    ASASSN-20kv                     & 2459088.2 & ---       & ---       & $-$3.25(7)    & 16.7(5)   & --- & --- \\
    ASASSN-20kw                     & 2459089.3 & $-$2.92(7) & 16.6(4)  & $-$6.8(1)    & 28.7(5)   & $-$14.3(1) & 33.5(2) \\
    MASTER J2139                    & 2459119.2 & 17.68(8)  & 13.67(6)  & 1.54(6)   & 6.3(3)   & --- & --- \\
    ASASSN-20mf                     & 2459119.2 & 7.69(3) &  8.69(3)  & ---    & ---   & --- & --- \\
\hline
\end{longtable}

\begin{figure*}[tb]
    \centering
    \includegraphics[width=15cm]{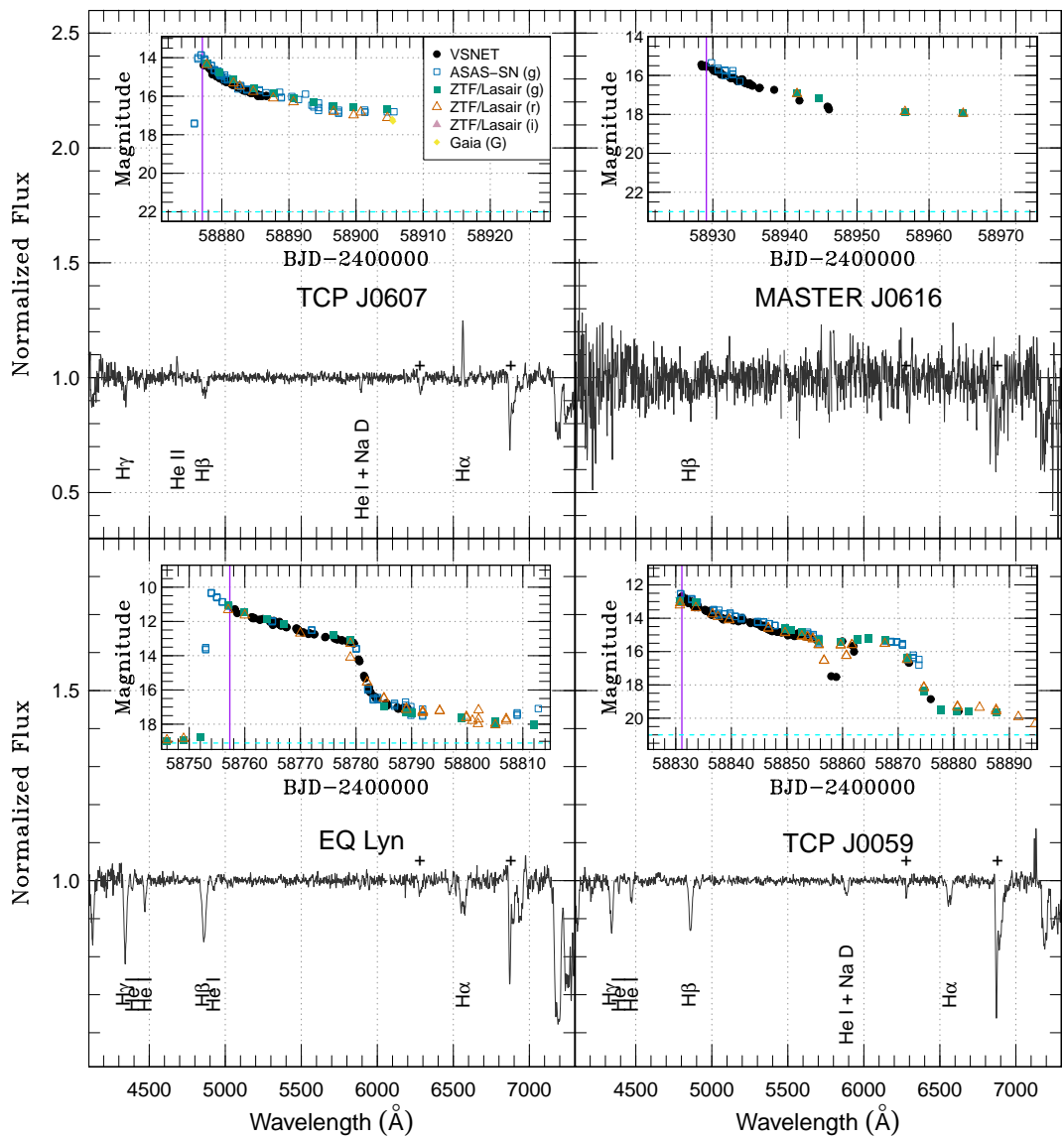}
    \caption
    {Normalized spectra and light curves during the outburst of TCP J0607 (upper left), MASTER J0616 (upper right), EQ Lyn (lower left) and TCP J0059 (lower right).
    The “+” signs in the spectra represent telluric absorptions.
    Solid line in the light curve figure represents the date of spectroscopic observations, and dashed line shows the quiescence or upper limit magnitude.
    Filled circles are data from VSNET \citep{VSNET}, filled squares, open triangles and filled triangles are the one from ZTF \citep{ZTF} or Lasair \citep{lasair} in the $g$, $r$ and $i$ band, open square are the one from ASAS-SN \citep{ASASSN} in the $g$ band, and filled diamonds are the one from Gaia alert \citep{wyr12gaiaalert} in the $G$ band, respectively.
    }
    \label{fig:lcspec1}
\end{figure*}

\begin{figure*}[tb]
    \begin{center}
            \includegraphics[width=15cm]{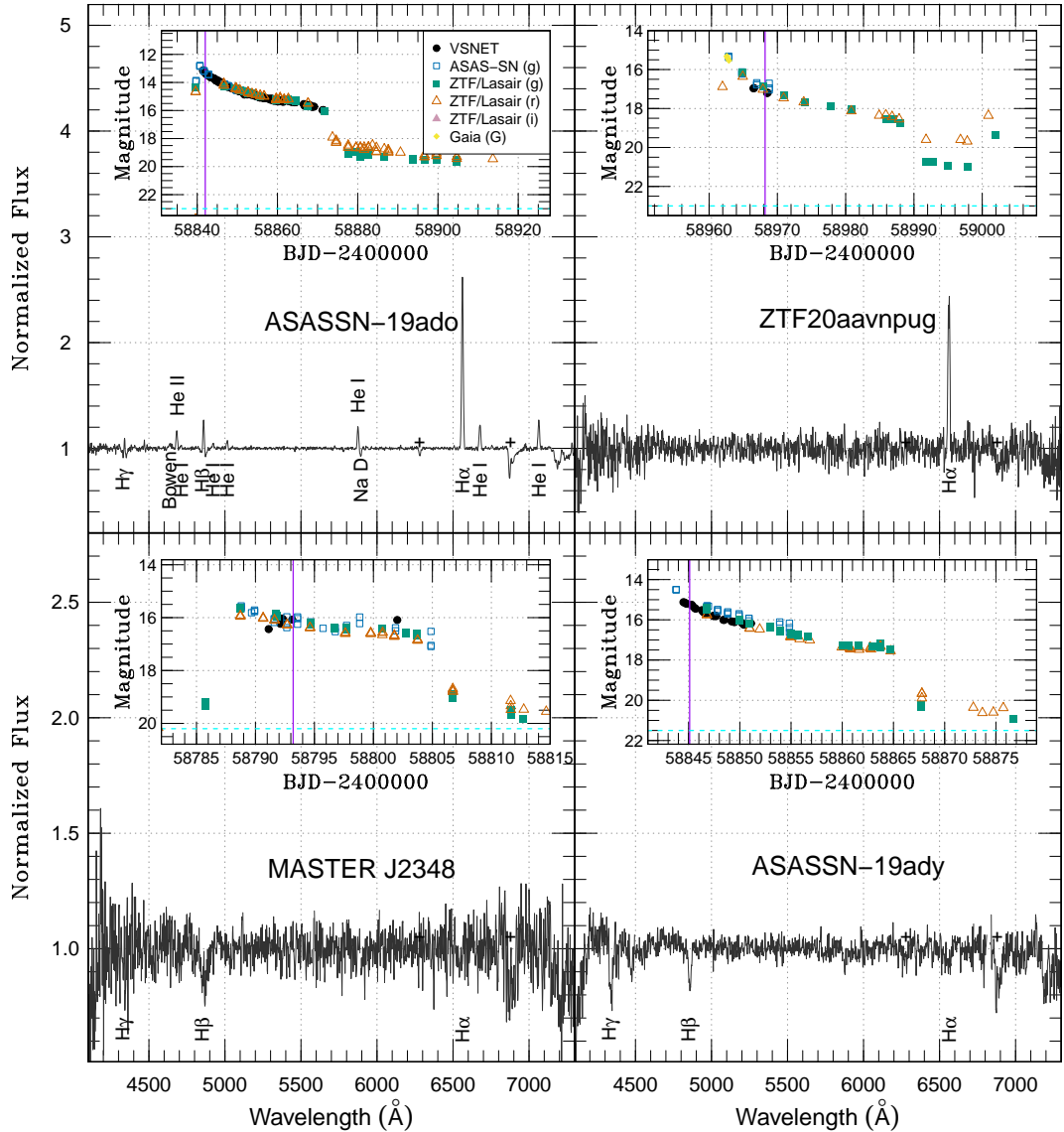}
    \end{center}
    \caption
    {Same figure as Figure \ref{fig:lcspec1} but of ASASSN-19ado (upper left), ZTF202aavnpug (upper right), MASTER J2348 (lower left) and ASASSN-19ady (lower right).
    }
    \label{fig:lcspec2}
\end{figure*}

\begin{figure*}[tb]
    \begin{center}
            \includegraphics[width=15cm]{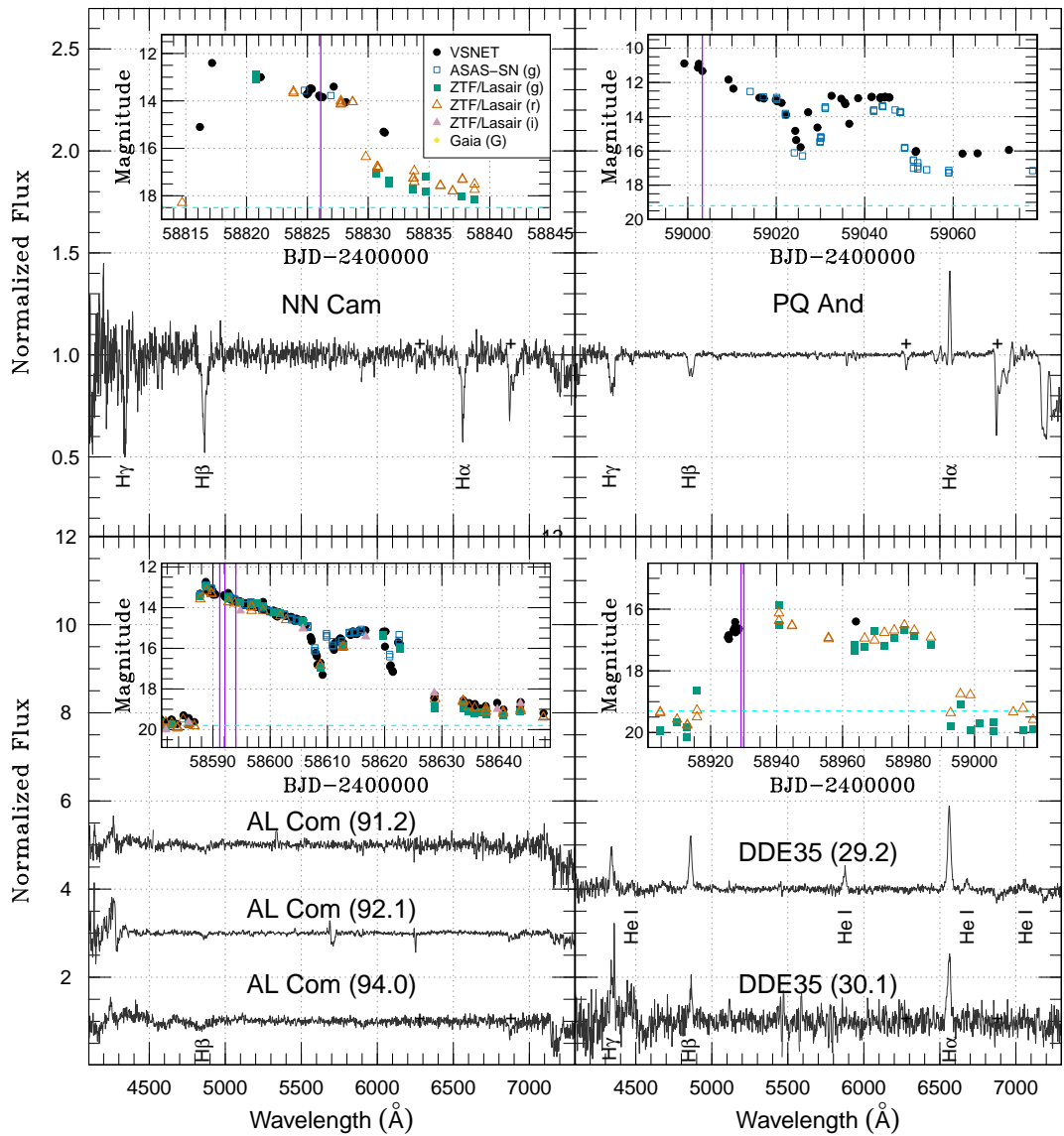}
    \end{center}
    \caption
    {Same figure as Figure \ref{fig:lcspec1} but of NN Cam (upper left), PQ And (upper right), AL Com (lower left) and DDE35 (lower right).
    }
    \label{fig:lcspec3}
\end{figure*}

\begin{figure*}[tb]
    \begin{center}
            \includegraphics[width=15cm]{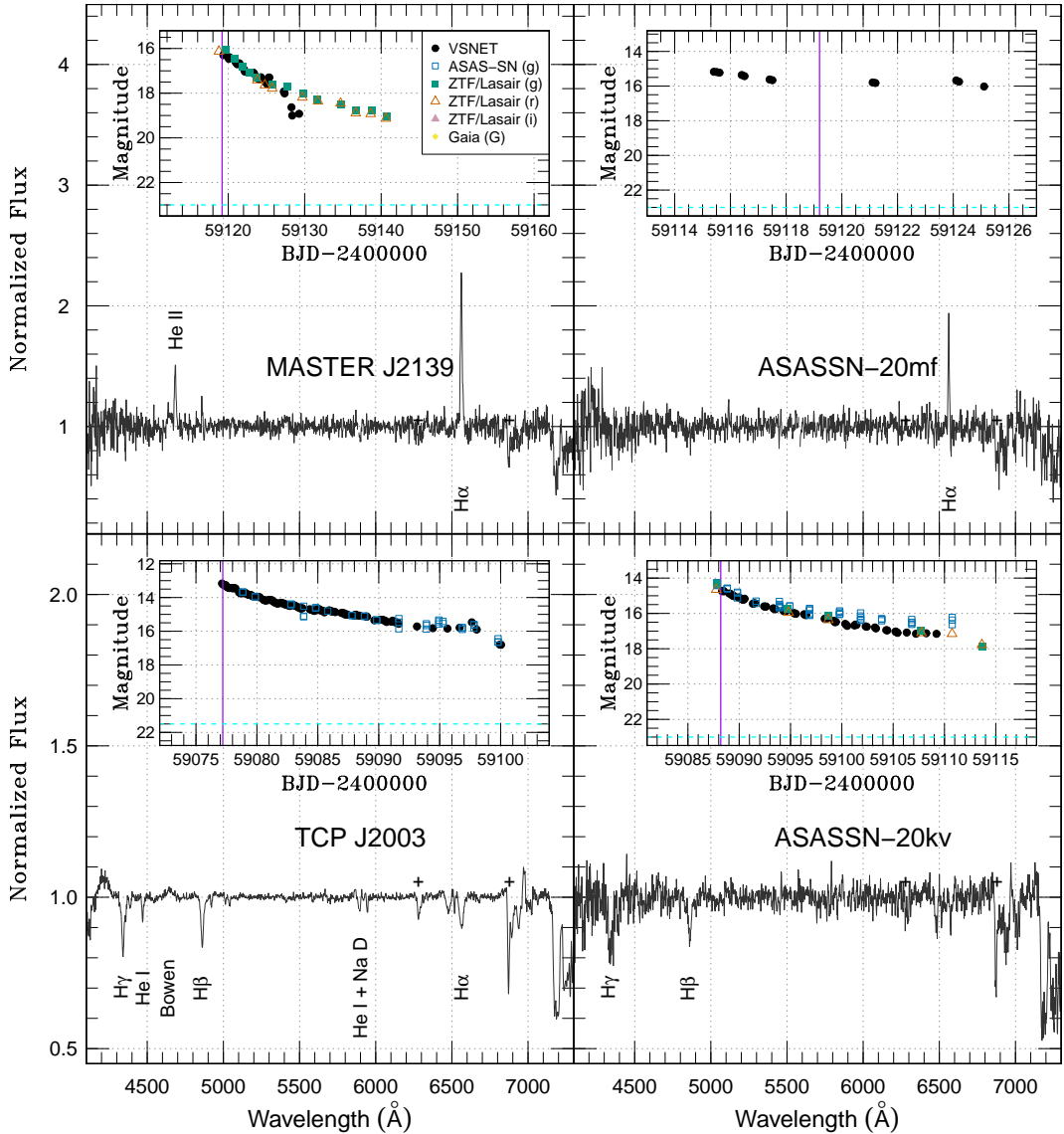}
    \end{center}
    \caption
    {Same figure as Figure \ref{fig:lcspec1} but of MASTER J2139 (upper left), ASASSN-20mf (upper right), TCP J2003 (lower left) and ASASSN-20kv (lower right).
    }
    \label{fig:lcspec4}
\end{figure*}

\begin{figure}[tb]
    \begin{center}
            \includegraphics[width=8cm]{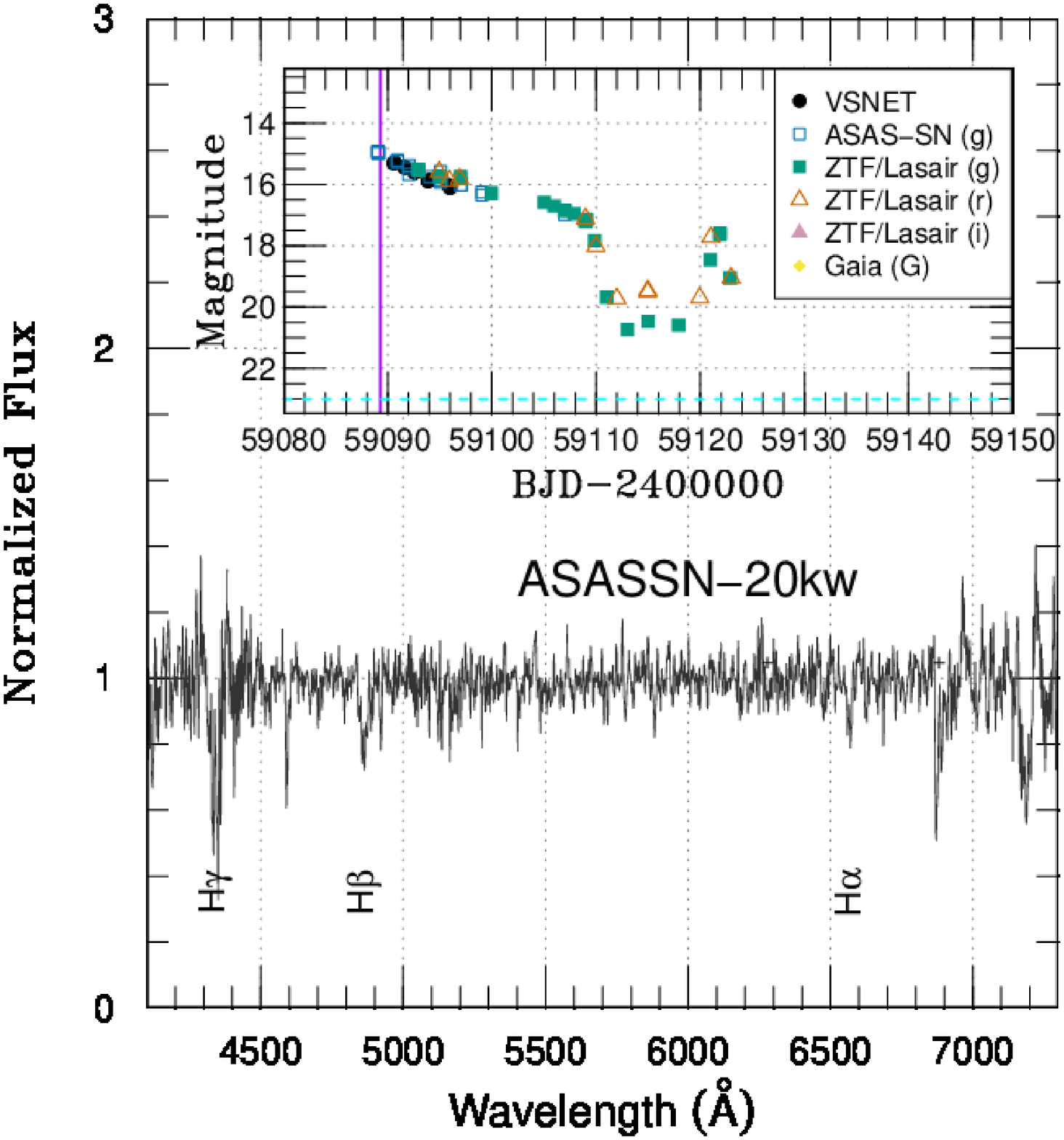}
    \end{center}
    \caption
    {Same figure as Figure \ref{fig:lcspec1} but of ASASSN-20kw.
    }
    \label{fig:lcspec5}
\end{figure}

We show our spectra and light curves of 17 sources in Figure \ref{fig:lcspec1}, \ref{fig:lcspec2}, \ref{fig:lcspec3}, \ref{fig:lcspec4} and  \ref{fig:lcspec5}.
In Table \ref{tab:2}, the full width at half maximum (FWHM) and equivalent width (EW) of Balmer lines of our spectra are listed.
In the light curves of Figure \ref{fig:lcspec1}-\ref{fig:lcspec5}, the quiescence magnitude or upper limit is presented as a dashed line, as well as  the the epoch when the spectroscopic observation was performed is shown with a solid line.
As seen in the light curves of Figure \ref{fig:lcspec1}-\ref{fig:lcspec5}, all our spectra were taken during their superoutbursts.
Among our 17 sources, 11 objects are WZ Sge-type DNe and their candidates based on the presence of early superhumps or global light curve profiles including rebrightenings.
The other six objects are SU UMa-type DNe and candidates.
Our spectra for six objects were obtained less than 1 d from the optical peak of their superoutbursts.

\begin{figure}[tbp]
 \begin{center}
    \includegraphics[width=80mm]{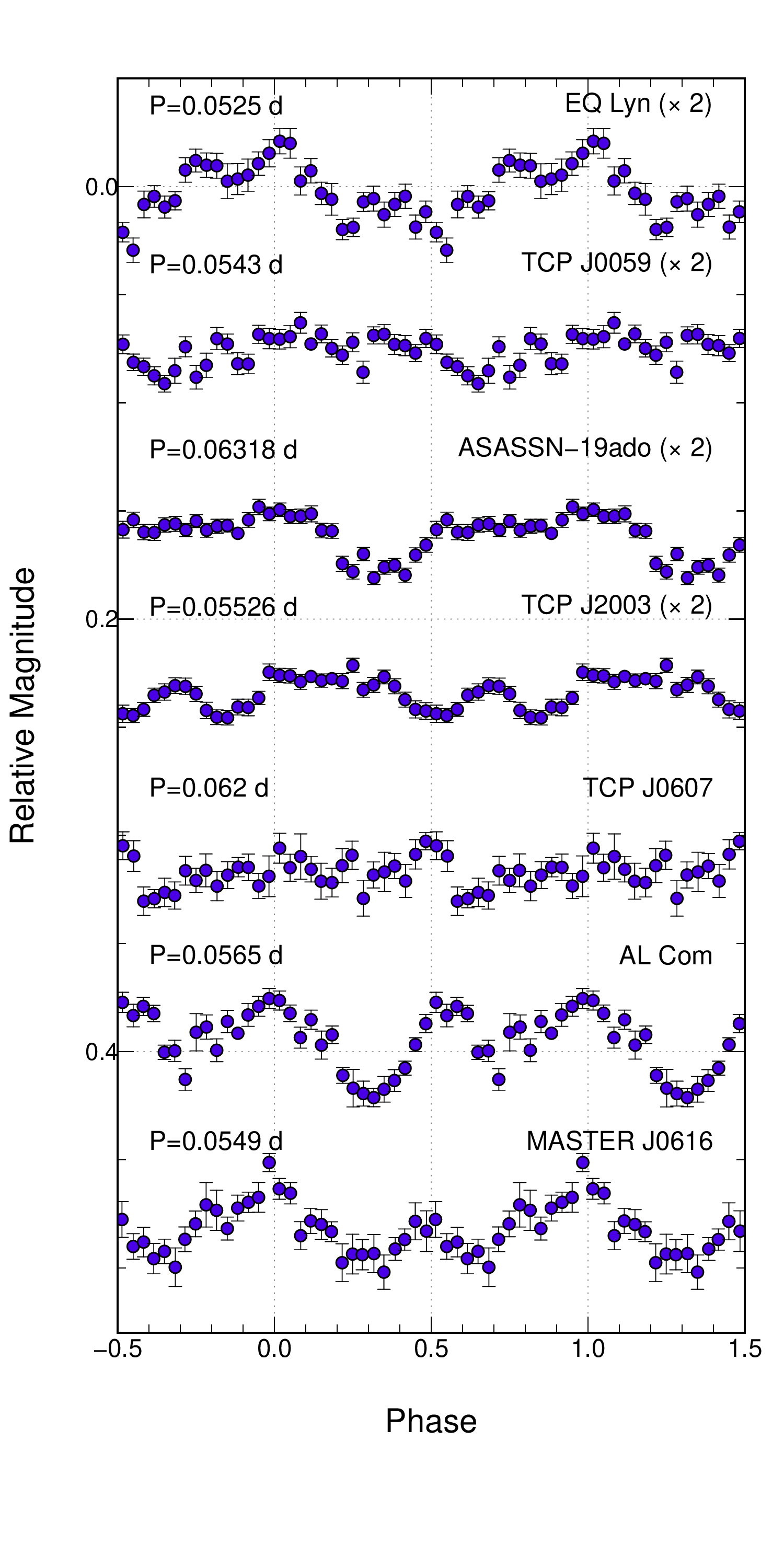}
  \end{center}
 \caption{Mean profiles of early superhumps of EQ Lyn, TCP J0059, ASASSN-19aod, TCP J2003, TCP J0607, AL Com, and MASTER J0616 from upper to bottom. 
 For the initial four objects, their magnitude scales are doubled for visualization.}
 \label{fig:earlyshs}
\end{figure}

\subsection{AL Com}
\label{sec:alcom}

AL Com is one of the well-studied WZ Sge-type DNe, with the orbital period of 0.056668 d \citep{Pdot}.
Its WZ Sge-type nature was firstly reported by \citet{kat96alcom}, and recent superoutbursts were observed in 2001 May \citep{ish02wzsgeletter}, 2007 October \citep{uem08alcom}, 2013 December \citep{Pdot6}, and 2015 March \citep{kim16alcom}, establishing the superoutburst cycle to be 6-7 yr.
The 2015 superoutburst was less energetic than other superoutbursts; its outburst amplitude was smaller than the other superoutbursts of AL Com and early superhumps were not observed.
These points suggest that the disk radius did not reach the 2:1 resonance radius of the system during the 2015 superoutburst \citep{kim16alcom}.

\begin{figure}[tbp]
 \begin{center}
    \includegraphics[width=85mm]{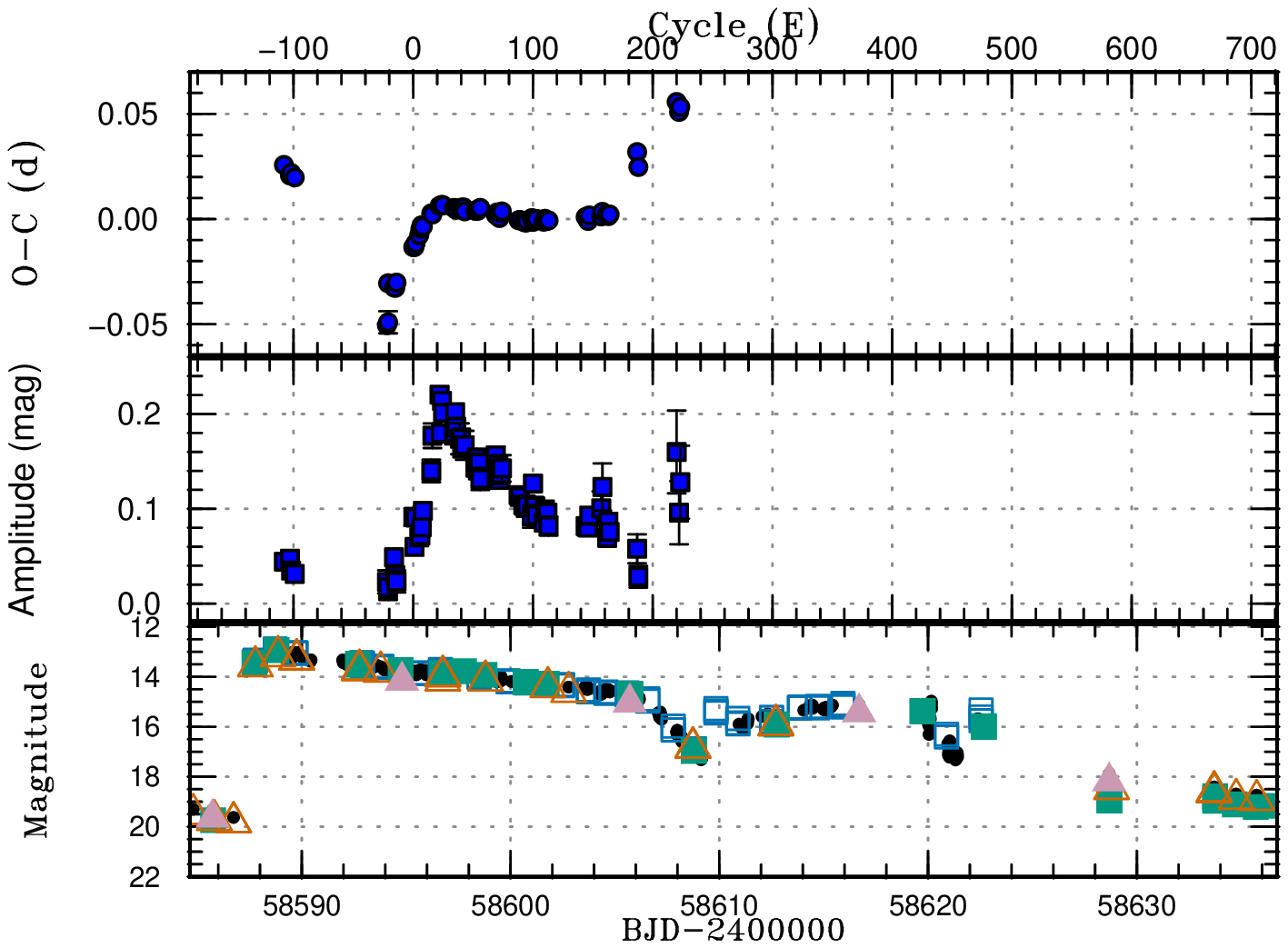}
  \end{center}
 \caption{ Top panel : the $O - C$ diagram of ALCom. 
    Note that $C =$ 0.057318 d was used to draw this figure following \citet{kim16alcom}.
    Times of superhump maxima are available in Table E21 in Supporting Information on the online version.
    Middle panel: the evolution of the superhump amplitudes of corresponding phase.
    Bottom panel: the light curve during the superoutburst.
    The symbols are same as Figure \ref{fig:lcspec1}.}
 \label{fig:alcomoc}
\end{figure}

The 2019 outburst was firstly reported by M. Moriyama on BJD 2458588.2 (vsnet alert 23171\footnote{http://ooruri.kusastro.kyoto-u.ac.jp/mailarchive/vsnet-alert/23171}).
This is only 4 yr after the last superoutburst in 2015.
The subsequent photometric observations detected early superhumps, confirming that the 2019 outburst is a WZ Sge-type superoutburst (Figure \ref{fig:earlyshs}). 
The period of early superhumps is 0.0565(1) d (Figure \ref{fig:earlyshs}), which is consistent with \citet{kat96alcom, ish02wzsgeletter, Pdot6}.
Thus the 2015 superoutburst had no impact on the 6-7 yr superoutburst cycle of AL Com.
The $O - C$ diagram of the superhump period and the evolution of superhump amplitude during the 2019 superoutburst are shown in Figure \ref{fig:alcomoc}.
The spectroscopic observations were performed on BJD 2458591.2, 2458592.1, and 2458594.2. 
Even though their  signal-to-noise ratios are low, H$\beta$ absorption line is present in all the three spectra (lower left panel of Figure \ref{fig:lcspec3}).
In our spectra, H$\alpha$ was not detected neither as emission nor absorption.
\citet{aug95alcomiauc} observed the 1995 superoutburst of AL Com 5 d after the outburst peak.
They reported the shallow and wide absorption lines of Balmer series and He I 4471\AA~, which is  likely consistent with our spectra.
The spectra during quiescence were studied by \citet{how98alcom, szk98alcom, szk03alcom}, in which AL Com showed the emission lines of Balmer series.

\subsection{EQ Lyn}
\label{sec:eqlyn}

\begin{figure}[tbp]
 \begin{center}
    \includegraphics[width=85mm]{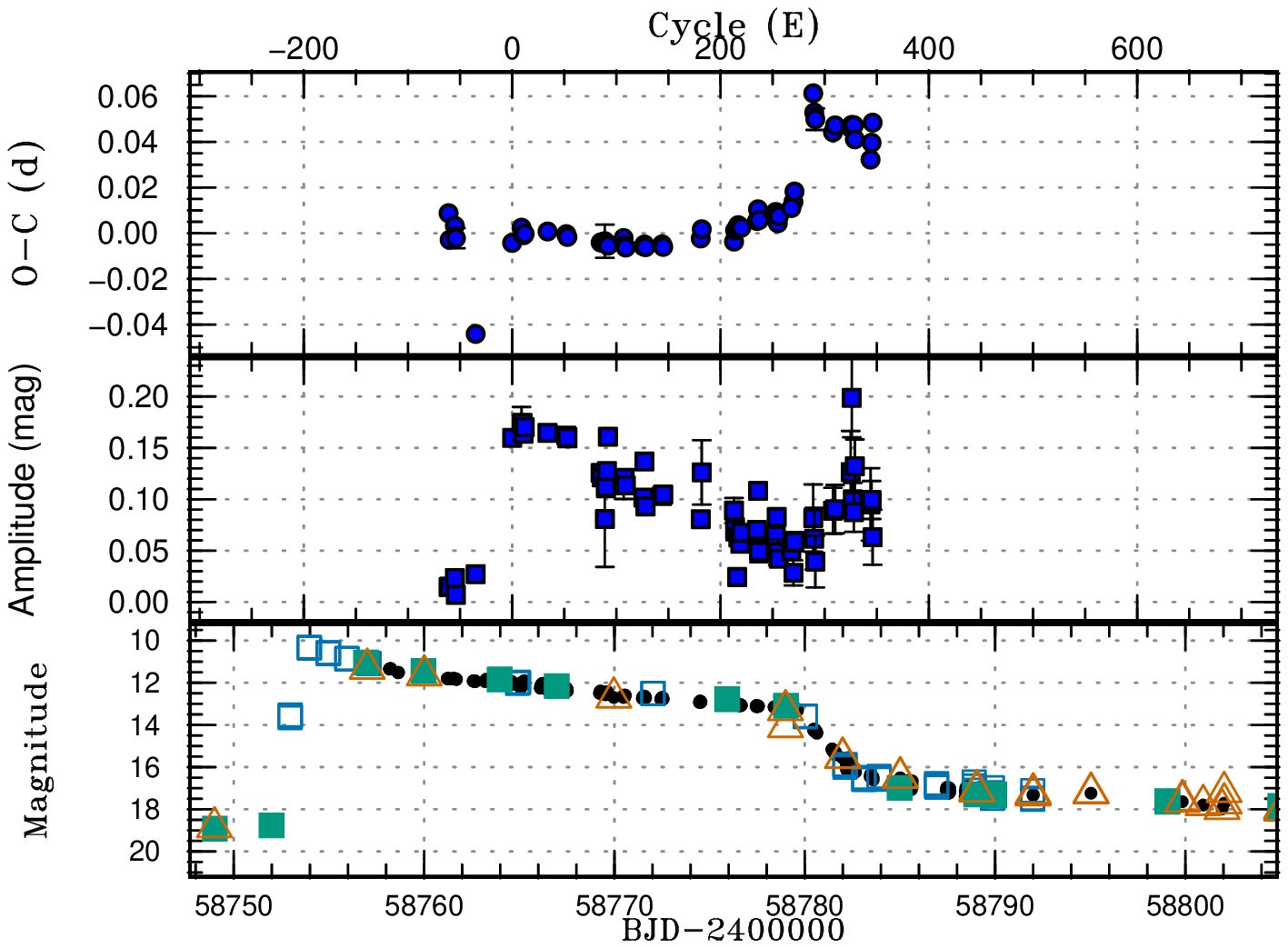}
  \end{center}
 \caption{ Top panel : the $O - C$ diagram of EQ Lyn. 
    Note that $C =$ 0.0547 d was used to draw this figure. 
    Times of superhump maxima are available in Table E19 in Supporting Information on the online version.
    Middle panel: the evolution of the superhump amplitudes of corresponding phase.
    Bottom panel: the light curve during the superoutburst.
    The symbols are same as Figure \ref{fig:lcspec1}.}
 \label{fig:eqlynoc}
\end{figure}

EQ Lyn (= SDSS J074531.92+453829.6; \cite{szk06SDSSCV5}) is a known WZ Sge-type DN with the orbital period of 0.0528(1) d \citep{gan09SDSSCVs}.
Its previous outburst in 2006 was recorded in CRTS \citep{CRTS}. 
The 2019 superoutburst  was firstly reported by T. Kojima on BJD 2458756.3 (vsnet alert 23591 \footnote{http://ooruri.kusastro.kyoto-u.ac.jp/mailarchive/vsnet-alert/23591}).
According to the ASAS-SN Sky Patrol data \citep{ASASSN}, EQ Lyn was in the superoutburst since BJD 2458753.0.
Our spectroscopic observation was performed on BJD 2458757.3 which is 3.3 d after the optical peak  \citep{iso19ateleqlyn}.  
Its spectra and  light curve during the 2019 superoutburst are presented in the lower left panel of Figure \ref{fig:lcspec1}.
The absorption lines of Balmer, He I 4388, 4471 and 4922 \AA~ are seen in the spectra, and a weak emission component is superposed in H$\alpha$.
Early superhumps are detected in our photometric observations with the period of 0.0525(1) d (Figure \ref{fig:earlyshs}).
No rebrightening was observed, hence EQ Lyn is a type-D object among WZ Sge-type DNe \citep{ima06tss0222,kat15wzsge}.
The $O - C$ diagram of the superhump period and the evolution of superhump amplitude during the 2019 superoutburst are shown in Figure \ref{fig:eqlynoc}.
The superhump period during Stage B was obtained as 0.054753(3) d.
The quiescence spectra from \citet{szk06SDSSCV5} show strong Balmer emission lines from the accretion disk and broader absorption lines from the primary WD.
Compared to its quiescence spectra, all Balmer lines in our spectra are turned into absorption, which is consistent with a spectrum of a DN outburst.

\subsection{MASTER OT J234843.23+250250.4}
\label{sec:j2348}

MASTER OT J234843.23+250250.4 (here after MASTER J2348) was firstly discovered in 2013 by \citet{shu13j2348atel5526}.
Photometric observations during the 2013 outburst confirmed that MASTER J2348 is a SU UMa-type DN with the superhump period of 0.032 d, which is below the period minimum \citep{Pdot6}. 
\citet{Pdot6} suggested that this object should be a Hydrogen-poor EI Psc-type DNe based on this superhump period, as Hydrogen-deficient AM CVn-type DNe with this $P_{\rm orb}$ do not show outbursts \citep{ram12amcvnLC}.

Our spectroscopic observation was performed on BJD 2458793.2 during its 2019 superoutburst, and our result is presented in the lower left panel of Figure \ref{fig:lcspec2}.
This observation revealed a weak H$\beta$ absorption line, although no strong He lines are  present \citep{iso19atelj2348}.
These features support that MASTER J2348 is an EI Psc-type DN rather than an AM CVn-type DN, as proposed by \citet{Pdot6}.

\subsection{NN Cam}
\label{sec:nncam}

\begin{figure}[tbp]
 \begin{center}
    \includegraphics[width=85mm]{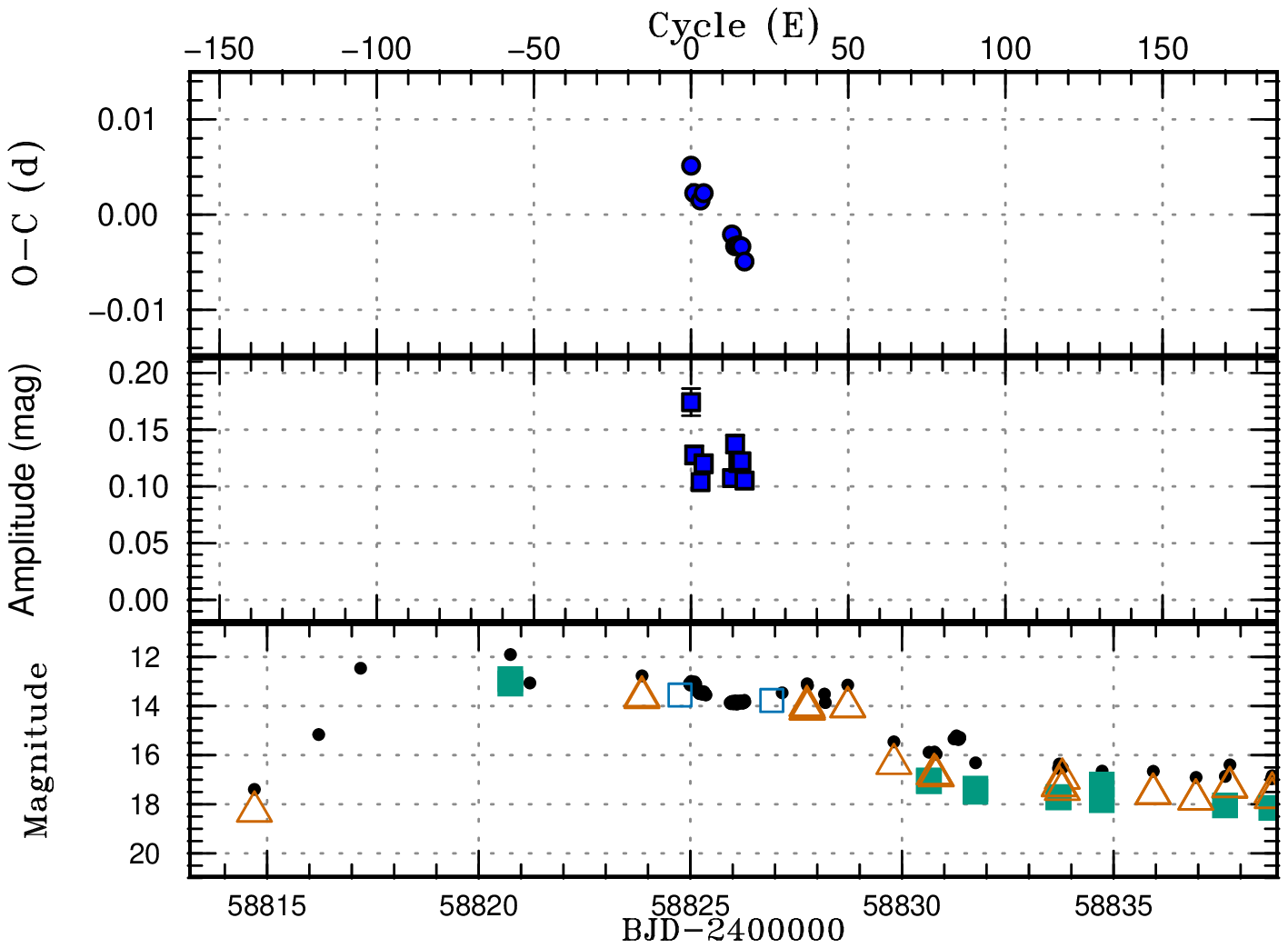}
  \end{center}
 \caption{ Top panel : the $O - C$ diagram of NN Cam. 
    Note that $C =$ 0.0743 d was used to draw this figure following \citet{Pdot7}. 
    Times of superhump maxima are available in Table E22 in Supporting Information on the online version.
    Middle panel: the evolution of the superhump amplitudes of corresponding phase.
    Bottom panel: the light curve during the superoutburst.
    The symbols are same as Figure \ref{fig:lcspec1}.}
 \label{fig:nncamoc}
\end{figure}

NN Cam is a known SU UMa-type DN with the orbital period of 0.0717 d, the superhump period of 0.07391(2) d, and the mass ratio estimated as 0.11 - 0.17 using the empirical relation of superhumps \citep{pat05SH}.
Its superoutbursts were studied in \citet{khr05nsv1485, Pdot, she11nncam}.
The 2019 outburst was firstly detected by Y. Maeda on BJD 2458816.2 and our spectroscopic observation was performed on BJD 2458826.1.
Our spectra in the upper left panel of Figure \ref{fig:lcspec3} show clear Balmer absorption lines, even though the peak epoch of this outburst is around BJD 2458817.2, which is 9 d prior to our spectroscopic observation.
Figure \ref{fig:nncamoc} shows the $O - C$ diagram during  the 2019 superoutburst, using the same $C$ = 0.0743 d as \citet{Pdot7}.
Judging from the shape of the $ O - C$ diagram, our time-resolved photometric observations were likely during Stage C, which also supports that our spectroscopic observation was in the late stage of the 2019 superoutburst.
These results suggest that NN Cam is a nearly face-on system, where the limb-darkening effect is relatively weak.

\subsection{TCP J00590972+3438357}
\label{sec:J0059}

\begin{figure}[tbp]
 \begin{center}
    \includegraphics[width=85mm]{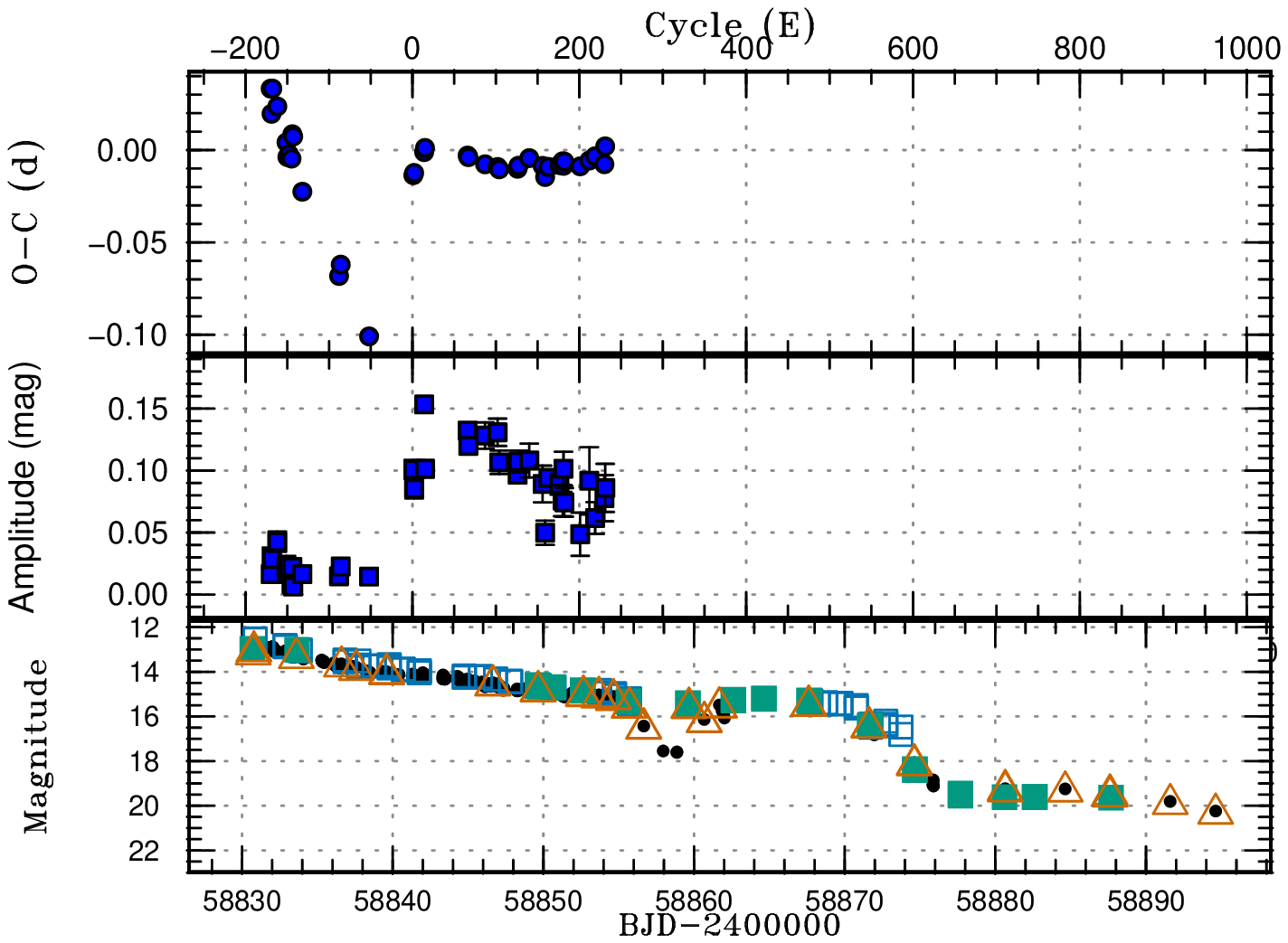}
  \end{center}
 \caption{ Top panel : the $O - C$ diagram of TCP J0059. 
    Note that $C =$ 0.0554 d was used to draw this figure. 
    Times of superhump maxima are available in Table E20 in Supporting Information on the online version.
    Middle panel: the evolution of the superhump amplitudes of corresponding phase.
    Bottom panel: the light curve during the superoutburst.
    The symbols are same as Figure \ref{fig:lcspec1}.}
 \label{fig:j0059oc}
\end{figure}

TCP J00590972+3438357 (= ASASSN-19adh = ZTF19acyfeuy, here after TCP J0059) was newly discovered by S. Ueda on BJD 2458830.9 at 12.4 mag \footnote{http://www.cbat.eps.harvard.edu/unconf/followups/J00590972+3438357.html}.
The quiescence counterpart is G = 21.073 mag from $\it{Gaia}$ DR2 \citep{GaiaDR2}, hence the outburst amplitude is $\sim$ 8.7 mag.
Our spectra were taken on BJD 2458831.1, which is only 0.2 d after the discovery  \citep{iso19atelj0059}.
The  spectra show Balmer,  He I 4388, 4471 and 5876\AA~ absorption lines, classifying this optical transient as a DN outburst (lower right panel of Figure \ref{fig:lcspec1}).
In H$\alpha$, a weak  emission component is superposed.
Possible Na D absorption line is seen as well.
From our photometric observations, we detect early superhumps with the period of 0.0543(1) d (Figure \ref{fig:earlyshs}), and Stage A superhump with the period of 0.0564(1) (Figure \ref{fig:j0059oc}).
These superhump periods yield that the mass ratio $q$ of TCP J0059 is 0.103(6) using the method proposed by \citet{kat13qfromstageA}.
TCP J0059 also showed a long-duration rebrightening (type-A rebrightening: \cite{ima06tss0222}).
Although the profile of early superhumps is not clearly double-wave due to its small amplitude, the long waiting time until the ordinary superhump appearance ($\sim 10$ d), short superhump period and rebrightening profile support that TCP J0059 is a WZ Sge-type DN. 

\subsection{ASASSN-19ado}
\label{sec:a19ado}

\begin{figure}[tbp]
 \begin{center}
    \includegraphics[width=85mm]{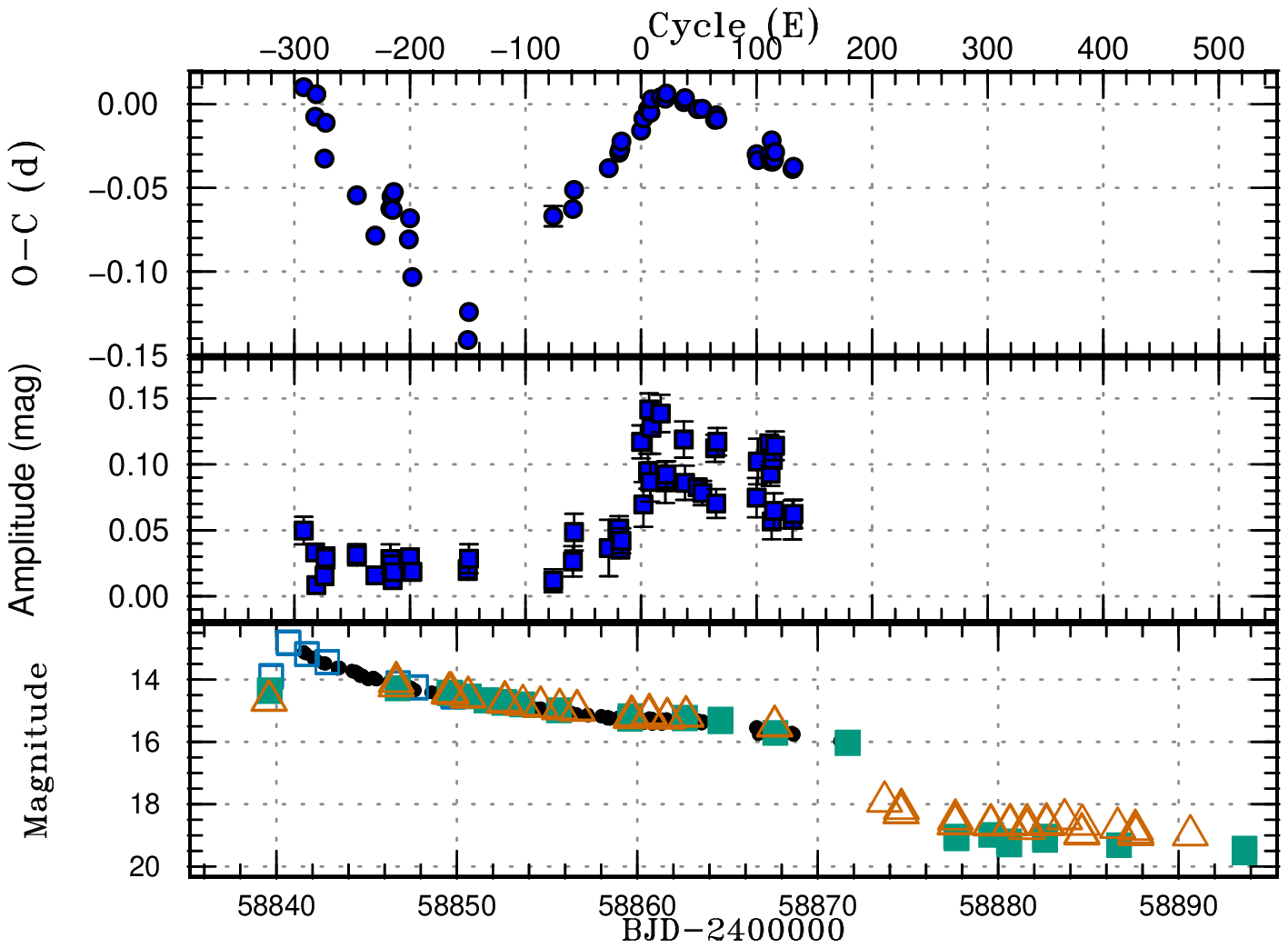}
  \end{center}
 \caption{ Top panel : the $O - C$ diagram of ASASSN-19ado. 
    Note that $C =$ 0.064085 d was used to draw this figure. 
    Times of superhump maxima are available in Table E23 in Supporting Information on the online version.
    Middle panel: the evolution of the superhump amplitudes of corresponding phase.
    Bottom panel: the light curve during the superoutburst.
    The symbols are same as Figure \ref{fig:lcspec1}.}
 \label{fig:a19adooc}
\end{figure}

ASASSN-19ado (= Gaia20abm = ZTF19adbgfwx) was firstly detected by ASAS-SN \citep{ASASSN} at 13.52 mag on BJD 2458839.7 as a CV candidate, and our spectroscopic observation was carried on BJD 2458842.0 \citep{mae19atela19ado}.
As the quiescence counterpart is likely $\sim$ 23 mag from PAN-STARRS1 image \citep{panstarrs1}, the outburst amplitude is $\sim$ 9 mag.
In the upper left panel of Figure \ref{fig:lcspec2},  the spectra show strong emission lines of Balmer and He I 4713, 4922, 5015, 5876, 6678, and 7065 \AA.
Na D absorption, relatively strong He II 4686 \AA~emission and  Bowen blend emission lines are visible in our spectra as well.
Our photometric observations detect early superhumps with the period of 0.06316(1) d (Figure \ref{fig:earlyshs}), establishing that ASASSN-19ado is a WZ Sge-type DN.
No rebrightening was observed, classifying as a type-D rebrightening object \citep{ima06tss0222}.
As the period of Stage A superhump was detected as 0.0651(1) d (Figure \ref{fig:a19adooc}), the mass ratio is estimated to be 0.081(5) following \citet{kat13qfromstageA}.

\subsection{ASASSN-19ady}
\label{sec:a19ady}

ASASSN-19ady was firstly discovered by ASAS-SN \citep{ASASSN} on BJD 2458843.8 at 14.42 mag as a CV candidate.
The optical quiescence counterpart is 21.5 mag in the $r$ band of Pan-STARRS1 catalog \citep{panstarrs1}, therefore the outburst amplitude is $\sim$ 7.1 mag. 
Before this 2019 outburst, no optical brightening was recorded by ASAS-SN \citep{ASASSN} nor ZTF \citep{ZTF} since 2013, which indicates the outburst cycle is likely longer than 6 yr.
The spectroscopic observation by Seimei telescope was performed on BJD 2458845.1 which is 1.3 d from the discovery \citep{iso19atela19ady}.
All Balmer lines are seen as absorption, which is consistent with the early phase spectra of a DN outburst (lower right panel of Figure \ref{fig:lcspec2}).
Although our photometric observations are not capable to confirm the presence of superhumps,
the large amplitude and long duration of outburst, and long outburst cycle suggest that ASASSN-19ady is a SU UMa-type DN candidate.

\subsection{TCP J06073081-0101501}
\label{sec:j0607}

TCP J06073081-0101501 (= ASASSN-20bk = Gaia20baq = ZTF20aakficj, here after TCP J0607) was found by Brazilian Transient Search on BJD 2458876.6 at 13.9 mag\footnote{http://www.cbat.eps.harvard.edu/unconf/followups/J06073081-0101501.html}.
Since the quiescence counterpart is not present in PAN-STARRS1 \citep{panstarrs1} image,  the outburst amplitude is likely larger than 8 mag. 
Our spectroscopic observation was performed on BJD 2458877.1, which is 0.5 d apart from the discovery \citep{mae20atelj0607}.
Our spectra in the upper left panel of Figure \ref{fig:lcspec1} show H$\alpha$ emission, and weak H$\beta$ and H$\gamma$ absorption lines. 
He II 4686 \AA~ emission and possible Na D absorption lines are present as well.
The subsequent photometric observations detected early superhumps with the period of 0.062(1) d (Figure \ref{fig:earlyshs}), therefore TCP J0607 is classified as a WZ Sge-type DN.
Due to the solar conjunction, its later light curve evolution was not observed.
Therefore its ordinary superhump periods and rebrightening profile are unknown.

\subsection{DDE35}
\label{sec:dde35}

DDE35 is a SU UMa-type DN candidate with the superhump period of 0.1322(1) d, which is above the period gap (vsnet-alert 24821 \footnote{http://ooruri.kusastro.kyoto-u.ac.jp/mailarchive/vsnet-outburst/24821}).
We note that this period is obtained using observations for only two nights and can be a spurious detection.
The spectra were obtained on BJD 2458929.2 and 2458930.1 (lower right panel of Figure \ref{fig:lcspec3}).
Both of our spectra show strong Balmer emission lines.
In the spectra on BJD 2458929.2,  He I 4471, 5876, 6678, and 7065 \AA~ are present as emission lines, which are less prominent on BJD 2458929.2 due to lower signal-to-noise ratio.
The spectra during the 2016 outburst reported by \citet{fra16atelDDE35} also showed narrow emission lines of Balmer series and He I 5876, 7065 \AA, which is consistent with our spectra in the 2020 superoutburst.
The quiescence spectra by \citet{wor17CVspec} show flat continuum and stronger Balmer and He I emission lines than in outburst.
Note that we cannot determine the peak epoch due to the lack of photometric observations and its peculiar double-peak outburst profile.
Judging from the light curve profile, our spectra were likely obtained in the first days of the outburst.

\subsection{MASTER OT J061642.05+435617.9}
\label{sec:j0616}

MASTER OT J061642.05+435617.9 (=ZTF20aaukvqw, here after MASTER J0616) was discovered by MASTER team on BJD 2458927.4 \citep{j0616_ateldiscovery}.
 \citet{j0616_ateldiscovery} reported that MASTER J0616 can be a classical nova.
As there is no VizieR \footnote{http://vizier.u-strasbg.fr/viz-bin/VizieR} match object, the outburst amplitude is larger than 8 magnitude.
Our spectra on BJD 2458929.1 in the upper right panel of Figure \ref{fig:lcspec1}, however, show narrow H$\beta$ absorption line with no clear sign of other lines \citep{iso20atelj0616}.
These features suggest this transient is a DN outburst in early phase rather than a nova eruption.
The subsequent photometric observations showed possible early superhumps with the period of 0.0549(1) d (Figure \ref{fig:earlyshs}).
Its large outburst amplitude and long waiting time until the superhump appearance also confirm that MASTER J0616 is a WZ Sge-type DN.

\subsection{ZTF20aavnpug}
\label{sec:z20aav}

ZTF20aavnpug (= Gaia20byj) was firstly detected by ZTF \citep{ZTF} at 16.9 mag on BJD 2458961.5 and Gaia alert \citep{wyr12gaiaalert} also reported as a new transient.
As there is no quiescence counterpart in Pan-STARRS1 \citep{panstarrs1} image, the outburst amplitude is likely larger than 7.5 mag.
There is no recorded outburst in ASAS-SN \citep{ASASSN}, ZTF \citep{ZTF} nor {\it  Gaia} alert \citep{wyr12gaiaalert}.
Our spectra on BJD 2458968.2 show strong H$\alpha$ emission line with no other clear features (upper right panel of Figure \ref{fig:lcspec2}: \cite{iso20atelz20aavnpug}).
Although our photometric observations were not capable to confirm superhumps, its large amplitude, steeper decline followed by exponential decay in the outburst profile, and rebrightening profile suggest that  ZTF20aavnpug is a WZ Sge-type DN candidate.

\subsection{PQ And}
\label{sec:pqand}

PQ And was discovered by D. McAdam in 1988 as a classical nova candidate, however later was classified as a DN due to the lack of nebular feature in the spectrum \citep{wad88pqandiauc}.
Moreover, \citet{ric90pqand} established its outburst cycle as $\sim$ 25 yrs from the archive data.
\citet{pat05pqand} found that its $P_{\rm orb} $ is $ \sim 0.0559(2)$ d.
These features established that PQ And is a WZ Sge-type DN.

Its outburst in 2020 was firstly reported by K. Hirosawa on BJD 2458998.2 (vsnet-alert 24301\footnote{http://ooruri.kusastro.kyoto-u.ac.jp/mailarchive/vsnet-alert/24301}), and our spectroscopic observation was carried out on BJD 2459003.3.
This was the first outburst observed since the initial discovery and outburst in 1988.  
This 2020 superoutburst was followed by a single rebrightening plus a long-duration rebrightening with a 2-mag dip during the rebrightening.
Such a dip during the long-duration rebrightening was observed in AL Com (Section \ref{sec:alcom}, \cite{kim16alcom}), ASASSN-15po \citep{nam16a15po}, V803 Cen (vsnet alert 19627\footnote{http://ooruri.kusastro.kyoto-u.ac.jp/mailarchive/vsnet-alert/19627}), and MASTER OT J172758.09+380021.5 (vsnet alert 23720\footnote{http://ooruri.kusastro.kyoto-u.ac.jp/mailarchive/vsnet-alert/23720}).
Therefore, a small dip might be a common feature in long-duration rebrightenings.
We note, since the 2020 superoutburst was discovered just after the solar conjunction, the peak of the superoutburst is likely before the discovery date.
Our spectra in the upper right panel of Figure \ref{fig:lcspec3}  show Balmer absorption lines and superposed emission lines in H$\alpha$ and H$\beta$ \citep{iso20atelpqand}.
Compared to the quiescence spectra reported in \citet{sch04pqand, zho20lamostDN}, absorption components of Balmer lines in our spectra are much stronger, which is consistent with the spectra during a DN outburst.
According to IAUC 4620 \citep{mca88pqandiauc}, the spectra  during  the 1988 outburst showed strong O III emission line, however our spectra in the 2020 outburst do not show any feature of O III. 
We suspect that, as PQ And was firstly thought to be a classical nova, they might mistake noise or cosmic ray for the O III emission line.
We note that as our spectra were taken in astronomical twilight and the elevation of the object was 19 degrees, those are strongly contaminated by telluric absorption and sunlight.

\subsection{TCP J20034647+1335125}
\label{sec:j2003}
\begin{figure}[tbp]
 \begin{center}
    \includegraphics[width=85mm]{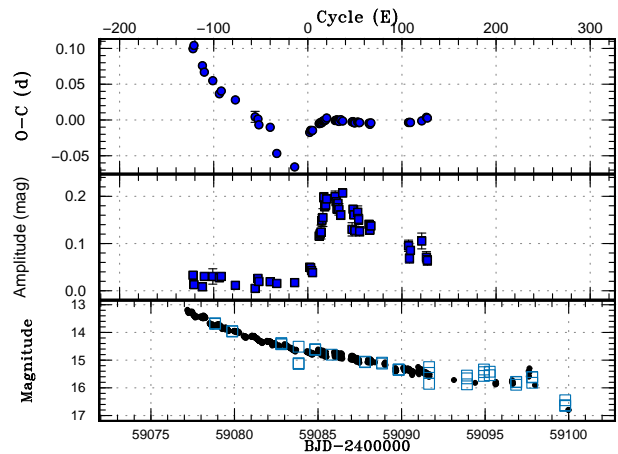}
  \end{center}
 \caption{ Top panel : the $O - C$ diagram of TCP J2003. 
    Note that $C =$ 0.0565 d was used to draw this figure. 
    Times of superhump maxima are available in Table E24 in Supporting Information on the online version.
    Middle panel: the evolution of the superhump amplitudes of corresponding phase.
    Bottom panel: the light curve during the superoutburst.
    The symbols are same as Figure \ref{fig:lcspec1}.}
 \label{fig:j2003oc}
\end{figure}

TCP J20034647+1335125 (=ASASSN-20kc, here after TCP J2003) was discovered on BJD 2459077.0 by H. Nishimura at 12.6 mag, which was not visible on 1 day prior.\footnote{http://www.cbat.eps.harvard.edu/unconf/followups/J20034647+1335125.html}
The quiescence counterpart is  21.5 mag  in the $r$ band from Pan-STARRS1 \citep{panstarrs1}, therefore the outburst amplitude of TCP J2003 is $\sim$ 8.9 mag.
Our spectroscopic observation was performed on BJD 2459077.2, which is only 0.2 d from the discovery \citep{tag20j2003atel}.
Our spectra are shown in the lower left panel of Figure \ref{fig:lcspec4},
which show Balmer, He I 4471 and 5876\AA~ absorption lines.
Possible Na D absorption and Bowen Blend emission lines are detected in our spectra as well.
These features confirmed TCP J2003 as a DN outburst. 
Our photometric follow up observations detected early superhumps with a period of 0.05526(4) d (Figure \ref{fig:earlyshs}), and Stage A superhump with a period of 0.05750(4) d (Figure \ref{fig:j2003oc}), yielding the mass ratio of TCP J2003 as  0.109(5) \citep{kat13qfromstageA}.
Therefore, TCP J2003 is classified as a WZ Sge-type DN, though the rebrightening profile is unknown.

\subsection{ASASSN-20kv}
\label{sec:a20kv}
\begin{figure}[tbp]
 \begin{center}
    \includegraphics[width=85mm]{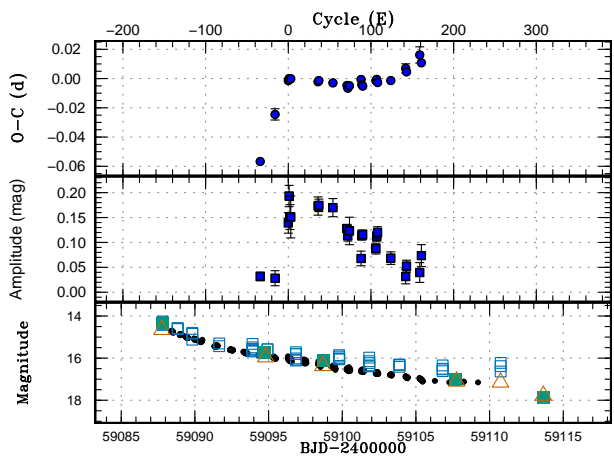}
  \end{center}
 \caption{ Top panel : the $O - C$ diagram of ASASSN-20kv. 
    Note that $C =$ 0.0560 d was used to draw this figure. 
    Times of superhump maxima are available in Table E25 in Supporting Information on the online version.
    Middle panel: the evolution of the superhump amplitudes of corresponding phase.
    Bottom panel: the light curve during the superoutburst.
    The symbols are same as Figure \ref{fig:lcspec1}.}
 \label{fig:a20kvoc}
\end{figure}

ASASSN-20kv (= Gaia20eeh = ZTF20abwvhwm) was discovered on BJD 2459087.8 at 14.1 mag in the $g$ band and reported as a possible classical nova in our Galaxy by ASAS-SN \citep{ASASSN}.
The quiescent counterpart is likely a blue faint object on Pan-STARRS1 image \citep{panstarrs1}, thus the outburst amplitude is $\sim$ 9 mag.
We observed ASASSN-20kv with Seimei telescope on BJD 2459088.2, which is 0.4 d from the initial detection by ASAS-SN.
The spectra in the lower right panel of Figure \ref{fig:lcspec4}, however, show weak H$\alpha$ and H$\beta$ absorption, revealing that ASASSN-20kv is a DN outburst rather than a classical nova eruption \citep{tag20a20kvatel}.
Our photometric observations detected the superhumps with a period of 0.05604(2) d during the Stage B (Figure \ref{fig:a20kvoc}).
The long waiting time until the superhump appearance ($\sim$ 7 d), its short superhump period and large superoutburst amplitude suggest that ASASSN-20kv is a WZ Sge-type DN.
We note that its rebrightening profile is unknown.

\subsection{ASASSN-20kw}
\label{sec:a20kw}
ASASSN-20kw (= Gaia20edl = ZTF20abxxzmk) was reported by ASAS-SN \citep{ASASSN} on BJD 2459089.1 at 14.7 mag in the $g$ band as a candidate of a galactic classical nova.
As the quiescent counterpart is likely a blue faint object on Pan-STARRS1 image \citep{panstarrs1}, the outburst amplitude is larger than 8 mag.
In our spectra on BJD 2459089.3, weak H$\beta$ and possible other Balmer absorption lines are detected (Figure \ref{fig:lcspec5}).
We, thus, classified ASASSN-20kw as a DN outburst, not as an eruption from a classical nova \citep{tag20a20kwatel}.
Our photometric observations were not capable to detect solid superhumps.
As ASASSN-20kw showed a single rebrightening and its outburst amplitude is larger than 8 mag, ASASSN-20kw should be a SU UMa-type DN, and even can be a WZ Sge-type DN.

\subsection{MASTER OT J213908.79+161240.2}
\label{sec:j2139}
\begin{figure}[tbp]
 \begin{center}
    \includegraphics[width=85mm]{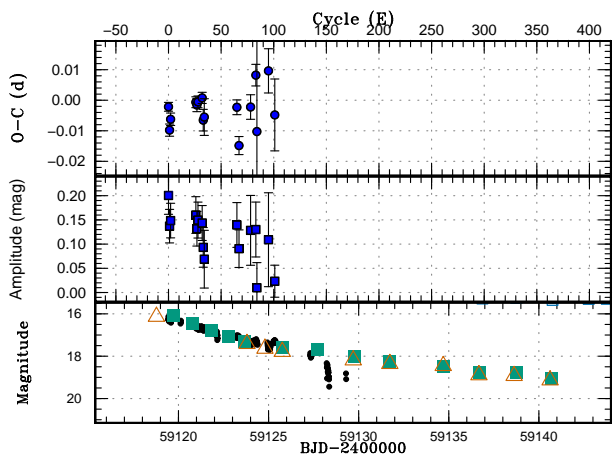}
  \end{center}
 \caption{ Top panel : the $O - C$ diagram of MASTER J2139. 
    Note that $C =$ 0.0584 d was used to draw this figure. 
    Times of superhump maxima are available in Table E26 in Supporting Information on the online version.
    Middle panel: the evolution of the superhump amplitudes of corresponding phase.
    Bottom panel: the light curve of MASTER J2139 during the superoutburst.}
 \label{fig:j2139oc}
\end{figure}

MASTER OT J213908.79+161240.2 (= ZTF20acerdbr, here after MASTER J2139) was firstly detected by MASTER-OAFA auto-detection system \citep{MASTER} on BJD 2459118.5 at 16.1 mag \citep{pog20j2139discoveryatel}.
As no quiescent counterpart is present on Pan-STARRS1 image \citep{panstarrs1},  the outburst amplitude is larger than 7 mag.
In the upper left panel of Figure \ref{fig:lcspec4}, our spectra on BJD 2459119.2 show strong emission lines of H$\alpha$ and He II 4686 \AA~ \citep{iso20j2139-a20mfatel}.
Such a strong emission line of He II 4686 \AA~ was observed in V455 And as well (\cite{nog09v455andspecproc}, Tampo et al. in prep). 
Our photometric observations detected a variation with the period of 0.0584(1) d and with the amplitude of $\sim$ 0.15 mag after 1 day from the discovery,
although it is not clear whether this was double-peaked or not due to low signal-to-noise ratio.
The $O - C$ diagram of the superoutburst is presented in Figure \ref{fig:j2139oc}.
The stable period and decreasing amplitude of superhumps are common in early superhumps.
Since V455 And also showed large-amplitude early superhumps with the amplitude of $\sim 0.2$ mag  \citep{kat09v455and}, the detected variations of MASTER J2139 can be early superhumps as well.
Considering its short superhump period and large outburst amplitude, MASTER J2139 would be a good candidate for a WZ Sge-type DN.

\subsection{ASASSN-20mf}
\label{sec:a20mf}
ASAS-SN \citep{ASASSN} detected ASASSN-20mf on BJD 2459114.9 at 14.5 mag in the $g$ band as a bright CV candidate. 
No quiescent counterpart is present on Pan-STARRS1 image \citep{panstarrs1}, which indicates that the outburst amplitude is larger than 8 mag.
We observed this object on BJD 2459119.2 with Seimei telescope, and our spectra in the upper right panel of Figure \ref{fig:lcspec4} show strong H$\alpha$ emission line \citep{iso20j2139-a20mfatel}.
Unfortunately, as ASAS-SN data was contaminated by a nearby star and there was no data in ZTF and Lasair, the global light curve profile is unknown.
Our photometric observations show a possible superhump, however are not enough to establish its period and amplitude.
As the outburst amplitude is larger than 8 mag and no past outburst is recorded in ASAS-SN nor ZTF, ASASSN-20mf is likely a SU UMa-type DN.

\section{Discussion}
\label{sec:4}

\subsection{spectroscopic observation in the early phase of dwarf nova outbursts}
\label{sec:4.1}

We report the spectra of 17 DN superoutbursts obtained by KOOLS IFU mounted on Seimei telescope.
It is worth noting that as the outbursts from 11 objects   (TCP J0059, ASASSN-19ado, TCP J0607, ZTF20aavnpug, ASASSN-19ady, MASTER J0616, TCP J2003, ASASSN-20kv, ASASSN-20kw, ASASSN-20mf, and MASTER J2139) are firstly detected in observed history, our quick spectroscopic observations enabled early classification of transient types and conducting follow-up observations.
The spectroscopic observations for 13 sources were performed less than 5 d from the outburst maxima.
The spectra of TCP J0059, TCP J0607, TCP J2003, ASASSN-20kv, ASASSN-20kw, and MASTER J2139 were taken less than 1 d from the outburst peaks and discovery reports.
These results  demonstrate that Seimei telescope has the capability to conduct quick follow-up observations of unknown transients.

\subsection{photometric classification of our objects}
\label{sec:4.11}

Among our objects, we newly confirmed TCP J0059, ASASSN-19ado, TCP J0607, MASTER J0616 and TCP J2003 as WZ Sge-type DNe based on the detection of early superhumps.
Even though early superhumps were not confirmed, ZTF20aavnpug, ASASSN-20kv and MASTER J2139 are good candidates of WZ Sge-type DNe based on the global light curve profile and short-period superhumps.
ASASSN-19ady, ASASSN-20kw and ASASSN-20mf are candidates of SU UMa-type DNe because of their large outburst amplitudes and light curve profiles.
Following the method given by \citet{kat13qfromstageA}, we estimated the mass ratio of TCP J0059 as 0.103(6), one of ASASSN-19ado as 0.081(5) and one of TCP J2003 as 0.109(5), respectively.
Their orbital periods, estimated mass ratios and rebrightening profiles are consistent with the distribution on the evolutionary path of CVs and WZ Sge-type DNe \citep{kat15wzsge}.

\subsection{correlations between photometric and spectroscopic features}
\label{sec:4.2}

High excitation lines such as He II and Bowen blend are often observed among WZ Sge-type DNe during their early superhump phase
(e.g. WZ Sge; \cite{bab02wzsgeletter,ste01spiralwave,nog04wzsgespec}, 
V592 Her; \cite{men02v592her}, 
CSS 080304:090240+052501; \cite{djo08j0902atel1411}, 
OT J111217.4-353829; vsnet-alert 9782 \footnote{http://ooruri.kusastro.kyoto-u.ac.jp/mailarchive/vsnet-alert/9782}; \cite{Pdot}, 
V572 And; \cite{qui05tss0222atel658, ima07j0222proc}, 
V1838 Aql; vsnet-alert 15779\footnote{http://ooruri.kusastro.kyoto-u.ac.jp/mailarchive/vsnet-alert/15779}; \cite{Pdot6}, 
ASASN-14cl; \cite{tey14asassn14clatel6235}, 
PNV J17292916+0054043; vsnet-alert 17327\footnote{http://ooruri.kusastro.kyoto-u.ac.jp/mailarchive/vsnet-alert/17327}; \cite{Pdot7},
V455 And; \cite{nog09v455andspecproc}; Tampo et al. in prep, 
and GW Lib; \cite{hir09gwlib}), 
though also seen in some phase in SS Cyg \citep{cla84sscyg, hes84sscyg} or high-inclination SU UMa-type DNe \citep{vog82zcha, hon88zcha, wu01iyuma}.
Among our 17 sources, high excitation lines, including He II 4686\AA~and Bowen blend, are detected in ASASSN-19ado (He II and Bowen blend), TCP J0607 (He II), TCP J2003 (Bowen blend), and MASTER J2139 (He II).
Except ASASSN-19ado, the spectra of the other thee objects were taken within 1 d from the superoutburst peaks.
All these objects are  WZ Sge-type DNe or candidates, supporting that high excitation lines  tend to be seen in WZ Sge-type DNe.

Na D absorption or emission line can be detected in the post-superoutburst stage of WZ Sge-type DNe, and a possible correlation between Na D line and mass reservoir have been discussed in \citet{pat98egcnc, vanspa10gwlib,neu17j1222, tam20j2104}.
However, it is not clear whether this correlation can be applied to  Na D line around the outburst peak.
Even though KOOLS-IFU cannot fully separate He I 5876 \AA~ and Na D, their asymmetry shapes imply that Na D absorption line is visible 
in TCP J0059, TCP J0607 ASASSN-19ado and TCP J2003.
Among these objects, TCP J0059 showed a long-duration rebrightening (type-A: \cite{ima06tss0222}).
On the other hand, ASASSN-19ado did not show any rebrightening (type-D).
We cannot judge the presence of rebrightenings in TCP J0607 and TCP J2003 since their post-superoutburst stages were not observed sufficiently.
As the spectra of these objects in post-superoutburst stages are not obtained, we cannot examine the difference of spectra around the optical peak and in the post-superoutburst stage. 
Therefore, direct comparison of spectra around the optical peak and post-superoutburst stage will be vital to establish the relation between the rebrightening types and Na D feature around the optical peak.

\citet{ste01spiralwave, bab02wzsgeletter} suggested that the spiral structure in the accretion disk obtained from the Doppler tomography of He II 4686 \AA~ might associate with double-peak early superhumps. 
Early superhumps are believed to be an inclination-depend variation as modeled by \citet{osa02wzsgehump, uem12ESHrecon}, and a more highly-inclined system tends to show larger-amplitude early superhumps.
In our objects, He II 4686 \AA~ was clearly observed only in ASASSN-19ado, TCP J0607 and MASTER J2139.
These three objects showed large-amplitude early superhumps around the optical peaks, even though we cannot confirm that the early superhump profile of MASTER J2139 is double-peaked.
In WZ Sge and V455 And, strong He II 4686\AA~ emission line was detected around the optical peaks as well \citep{nog04wzsgespec, nog09v455andspecproc}.
These objects showed early superhumps with the amplitude larger than 0.1 mag \citep{Pdot}.
The inclinations of WZ Sge and V455 And are estimated as 75 $\pm$2 and 75 degree, respectively \citep{sma93wzsge, ara05v455and}.
Both the He II 4686\AA~ emission line and early superhumps are observed in CSS 080304:090240+052501, OT J111217.4-353829,  V572 And, ASASN-14cl, and PNV J17292916+0054043 as well.
The amplitudes of early superhumps of these objects are larger than 0.15 mag \citep{kat15wzsge}.
We note that, as only the detections of He II emission feature are reported on these objects, the strengths of emission are not noted in the literature.
On the other hand, even though low-amplitude early superhumps are detected in EQ Lyn, TCP J0059 and TCP J2003 from our sources, their spectra did not show detectable He II emission line. 
During the superoutburst of V592 Her, very weak He II 4686\AA~ emission line was detected around the outburst peak \citep{men02v592her}, which showed early superhumps with the amplitude of 0.01 mag \citep{Pdot2}.
ASASSN-20kv, which had a long waiting time until the ordinary superhump appearance likely reflecting the early superhump stage, did not show He II emission line as well.
In GW Lib, whose inclination is $\sim$ 11 degree \citep{tho02gwlibv844herdiuma}, even though early superhumps are not observed, very weak He II 4686\AA~ emission line is visible in its spectra around the optical peak \citep{hir09gwlib}.
These objects are summarized in Table \ref{tab:heii}.
Above discussion suggests that the strength of the He II 4686\AA~ emission line is likely bound to the amplitude of early superhumps, and therefore also related to the inclination of the system.
This result can be understood with the occurrence of tidal instability at the 2:1 resonance radius, which causes a spiral structure in an accretion disk \citep{lin79lowqdisk, osa02wzsgehump}.
The vertical deformation of the accretion disk results in a larger early superhump amplitude in more highly inclined systems \citep{osa02wzsgehump,uem12ESHrecon}.
At the same time, He II 4686\AA~line is emitted from this heated and/or irradiated spiral structure \citep{mor02DNspectralatlas}.
Because of foreshortening and limb-darkening effects, a system with higher inclination should show a stronger emission line. 
To determine the practical relation between  early superhumps and profiles of He II 4686 \AA~ emission line, more objects and comprehensive observations are needed.
We note that, non detections of He II emission line in MASTER J0616 and AL Com, which showed clear early superhumps, are likely due to the low signal-to-noise ratio of their spectra.

\begin{longtable}{cccc}
  \caption{Strength of He II 4686\AA~ and amplitude of early superhumps}\label{tab:heii}
  \hline              
    Name & strength of  & amplitude of  & reference of spectra \\ 
     & He II 4686\AA\commente & early superhumps &  \\ 
     & & (mag) & \\
\endfirsthead
  \hline
    \multicolumn{4}{l}{\commente Peak flux strength compared the normalized continuum.} \\
    \multicolumn{4}{l}{\commentc "observed" means  He II 4686\AA~ emission line was detected, however the strength is not noted.} \\
    \multicolumn{4}{l}{\commenta Taken from \citet{kat15wzsge}}\\
\endfoot
  \hline
  \hline
    ASASSN-20kv                 & $<$ 0.05           & 0.00      & this work\\
    GW Lib                      & 1.03               & 0.00\commenta  & \citet{hir09gwlib}\\
    V592 Her                    & 1.01               & 0.01\commenta  & \citet{men02v592her} \\
    TCP J2003                   & $<$ 0.01           & 0.012        & this work\\
    TCP J0059                   & $<$ 0.01           & 0.016        & this work\\
    ASASSN-19ado                & 1.16               & 0.016        & this work\\
    EQ Lyn                      & $<$ 0.01           & 0.025        & this work\\
    TCP J0607                   & 1.10               & 0.028        & this work\\
    WZ Sge                      & 1.17               & 0.14\commenta  & \citet{nog04wzsgespec}\\
    MASTER J2139                & 1.50               & 0.189        & this work\\
    V455 And                    & 2.60               & 0.22\commenta  & \citet{nog09v455andspecproc}\\
    PNV J17292916+0054043       & observed\commentc  & 0.015\commenta & vsnet-alert 17327\\
    ASASSN-14cl                 & observed\commentc  & 0.018\commenta & \citet{tey14asassn14clatel6235}\\
    V572 And                    & observed\commentc  & 0.07\commenta  & \citet{qui05tss0222atel658}\\
    OT J111217.4-353829         & observed\commentc  & 0.14\commenta  & vsnet-alert 9782\\
    CSS 080304:090240+052501    & observed\commentc  & 0.35\commenta  & \citet{djo08j0902atel1411}\\
\end{longtable}

Next, we discuss the dependence of Balmer line profiles on systems using our sources whose spectra were taken less than 1 day from the optical peak (TCP J0059, TCP J0607, TCP J2003, ASASSN-20kv and MASTER J2139).
Note that, though the spectra of ASASSN-20kw was taken within 0.2 day from the discovery, we excluded this object from the following discussion, as its photometric observations relatively lack.
In the well observed systems (e.g. \cite{nog04wzsgespec,hir09gwlib}), it is known that spectral profile changes over an outburst.
We therefore suppose the above selection criteria for more statistical constraints.
In our sources, TCP J0059, TCP J2003, and ASASSN-20kv showed all Balmer lines in absorption.
This feature suggests that these three objects are mid- or low-inclined systems, which is in agreement with the respect that they showed low-amplitude early superhumps or did not show them.
In ASASSN-14cl, \citet{tey14asassn14clatel6235}  detected all Balmer series in absorption 0.5 d after the discovery.
ASASSN-14cl showed early superhumps with the amplitude of 0.018 mag \citep{kat15wzsge}. 
By contrast, in our sources,  H$\alpha$ line is in emission in TCP J0607 and MASTER J2139.
As proposed based on their large-amplitude early superhumps, this emission profile of H$\alpha$ also suggest that they are high-inclination system.
In literature, WZ Sge, whose early superhump amplitude is 0.14 mag, showed H$\alpha$ in emission and other Balmer seiries in absorption \citep{nog04wzsgespec}.
CSS 080304:090240+052501 and V455 And showed all Balmer lines in emission in their spectra around the optical peak, whose early superhump amplitudes are 0.22 and 0.35 mag, respectively \citep{djo08j0902atel1411, nog09v455andspecproc, kat15wzsge}.
These discussions support the general spectroscopic behavior that a system with higher (lower) inclination tends to show larger (smaller) early superhump amplitude, and stronger (weaker) emission components in Balmer line profiles.
We note, one exception in literature is low-inclined GW Lib \citep{tho02gwlibv844herdiuma}.
Even though GW Lib did not show detectable early superhumps \citep{Pdot}, the spectra within 1 d from the optical peak showed strong H$\alpha$ emission \citep{hir09gwlib}.
However, H$\alpha$ of GW Lib turned into absorption 10 d later. 
Therefore, more systematic studies of the spectral evolution across a DN superoutburst are needed to define the relation between the line profiles and system parameters. %

\section{Summary}
\label{sec:5}

In this paper, we present 17 DN superoutbursts observed by the 3.8m telescope Seimei and the VSNET collaboration.
Our findings are summarized as follows.

\begin{itemize}

\item
Based on our photometric observations through VSNET and using public survey data, 11 of our objects are WZ Sge-type dwarf novae and their candidates, and the other six objects are SU UMa-type dwarf novae and their candidates.
11 systems are newly identified systems in this work.
Using the periods of early and ordinary superhumps, we determined the mass ratios of TCP J0059, ASASSN-19ado and TCP J2003 as 0.103(6), 0.081(5) and 0.109(5), respectively.

\item
Our spectroscopic observations of 13 objects are performed within 5 d from the outburst peak of their superoutbursts.
Moreover, our spectra of six objects are obtained less than 1 d since the outburst peaks and discovery reports.
These quick follow-up observations illustrate the capability of Seimei telescope for follow-up observations of unknown transients.  

\item
In our objects, the emission line of He II 4686\AA~ is detected among ASASSN-19ado, TCP J0607 and MASTER J2139, who show large-amplitude early superhumps.
Combining nine WZ Sge-type DNe whose early superhump amplitudes and spectra during the early superhump phase are available in previous studies,
our result suggests that the higher the inclination of system is, the stronger the emission line of He II 4686\AA~is, as with the amplitude of early superhumps.

\item
Using our sources and previously reported spectra of WZ Sge-type DNe taken within 1 d from the outburst peaks, we examined a general correlation among the spectra in the early phase of WZ Sge-type dwarf novae.
The supported correlation is that
a system with higher (lower) inclination tends to show larger (smaller) early superhump amplitude, and stronger (weaker) emission components in Balmer series. 

\item
AL Com showed an outburst in 2019, which is only 4 yr later from the less energetic SU UMa-type superoutburst in 2015.
Our photometric observations confirmed that this 2019 outburst is a WZ Sge-type DN superoutburst, based on the detection of early superhumps.

\item 
We reported the observations on the second outburst of PQ And in observed history.
Our observation detected the plateau phase of the main superoutburst and long-duration rebrightening with a 2-mag dip, supporting its classification as a WZ Sge-type DN.

\item
MASTER J2348 is confirmed as an EI Psc-type DN based on the detection of H$\beta$ absorption line, whose superhump period is below the period minimum.
This classification is consistent with the suggestion given by \citet{Pdot6}.

\end{itemize}

\begin{ack}

We thank world wide professional and amateur observers who have shared data with VSNET and VSOLJ.
K. Maeda acknowledges support from the Japan Society for the Promotion of Science (JSPS) KAKENHI grant JP18H05223, JP20H00174, and JP20H04737.
U. Burgaz acknowledges the support provided by the Turkish Scientific and Technical Research Council (T\"UB\.ITAK-2211C and 2214A).  
This work was supported by the project APVV-15-0458 “Interacting binaries - Key for the Understanding of the Universe”.
This work was supported by the Slovak Research and Development Agency
under the contract No. APVV-15-0458 and by the Slovak Academy of Sciences
grant VEGA No. 2/0030/21.
The authors from Sternberg institute thank
the Program of Development of Lomonosov MSU ‘Leading Scientific Schools’. A.A.B, N.P.I. and  M.A.B. (SAI MSU) are supported by the Interdisciplinary Scientific and Educational School of Moscow University 'Fundamental and Applied Space Research'.
This research was partly supported by the Slovak Academy of Sciences grant VEGA No. 2/0030/21.
We acknowledge ESA Gaia, DPAC and the Photometric Science Alerts Team (http://gsaweb.ast.cam.ac.uk/alerts).
Based on observations obtained with the Samuel Oschin 48-inch Telescope at the Palomar Observatory as part of the Zwicky Transient Facility project. ZTF is supported by the National Science Foundation under Grant No. AST-1440341 and a collaboration including Caltech, IPAC, the Weizmann Institute for Science, the Oskar Klein Center at Stockholm University, the University of Maryland, the University of Washington, Deutsches Elektronen-Synchrotron and Humboldt University, Los Alamos National Laboratories, the TANGO Consortium of Taiwan, the University of Wisconsin at Milwaukee, and Lawrence Berkeley National Laboratories. Operations are conducted by COO, IPAC, and UW.
Lasair is supported by the UKRI Science and Technology Facilities Council and is a collaboration between the University of Edinburgh (grant ST/N002512/1) and Queen’s University Belfast (grant ST/N002520/1) within the LSST:UK Science Consortium. This research has made use of ''Aladin sky atlas'' developed at CDS, Strasbourg Observatory, France 2000A\&AS..143...33B and 2014ASPC..485..277B.

\end{ack}

\section*{Supporting Information}
The following Supporting Information is available on the online version of this article; Table E1 - E26.


\bibliographystyle{pasjtest1}
\bibliography{cvs}


\end{document}